\RequirePackage{fix-cm}
\documentclass{CVM}

\CVMsetup{
type      = {Research/Review Article},
doi       = {s41095-0xx-xxxx-x},
title     = {NCP: Neighborhood-Preserving Non-Uniform Circle Packing for Visualization},
author    = {Duan Li$^{1}$, Jun Yuan$^{1}$, Xinyuan Guo$^{1}$, Xiting Wang$^{2}$, Yang Liu$^{3}$, Weikai Yang$^{4}$\cor{}, and Shixia Liu$^{1}$\\
},
runauthor = {D. Li, J. Yuan, X. Guo, X. Wang, Y. Liu, W. Yang, S. Liu},
abstract  = {
    Circle packing is widely used in visualization due to its aesthetic appeal and simplicity, particularly in tasks where the spatial arrangement and relationships between data are of interest, such as understanding proximity relationships (\eg, images with categories) or analyzing quantitative data (\eg, housing prices).
    Many applications require preserving neighborhood relationships while encoding a quantitative attribute using radii for data analysis.
    To meet these two requirements simultaneously, we present a \emph{neighborhood-preserving non-uniform circle packing} method, NCP.
    This method preserves neighborhood relationships between the data represented by non-uniform circles to comprehensively analyze similar data and an attribute of interest.
    We formulate neighborhood-preserving non-uniform circle packing as a planar graph embedding problem based on the circle packing theorem.
    This formulation leads to a non-convex optimization problem, which can be solved by the continuation method. 
    We conduct a quantitative evaluation and present two use cases to demonstrate that our NCP method can effectively generate non-uniform circle packing results.
},
keywords  = {Circle packing, neighborhood preservation,  graph embedding, power diagram, force-directed method},
copyright = {The Author(s)},
}
\usepackage{hyperref}
\usepackage[capitalize]{cleveref}
\usepackage{subcaption}
\usepackage{enumitem}

\def \etal {{\emph{et al}.\thinspace}}
\def \eg {{\emph{e.g}.\thinspace}}

\DeclareMathOperator*{\Area}{Area}
\DeclareMathOperator*{\Enve}{Env}
\DeclareMathOperator*{\Convexhull}{CH}

\begin{document}

\maketitle

    \begin{figure}[b] \vskip -2mm
    \small\renewcommand\arraystretch{1.3}
        \begin{tabular}{p{80.5mm}} \toprule\\ \end{tabular}
        \vskip -4.5mm \noindent \setlength{\tabcolsep}{1pt}
        \begin{tabular}{p{3.5mm}p{80mm}}
    $1\quad $ & School of Software, BNRist, Tsinghua University, 100084, Beijing, China. E-mail: \{ld23, yuanj19, guo-xy24\}@mails.tsinghua.edu.cn, shixia@tsinghua.edu.cn.\\
    $2\quad $ & Gaoling School of Artificial Intelligence, Renmin University, Beijing, China. E-mail: xitingwang@ruc.edu.cn.\\
     $3\quad $ & Internet Graphics Group, Microsoft Research Asia, Beijing, China. E-mail: yangliu@microsoft.com.\\
    $4\quad $ & Hong Kong University of Science and Technology (Guangzhou), 511453, Guangzhou, China. E-mail: weikaiyang@hkust-gz.edu.cn.\\
&\hspace{-5mm} Manuscript received: 2022-01-01; accepted: 2022-01-01\vspace{-2mm}
    \end{tabular} \vspace {-3mm}
    \end{figure}


\section{Introduction}

\begin{figure*}[t]
\centering
  \includegraphics[width=\linewidth]{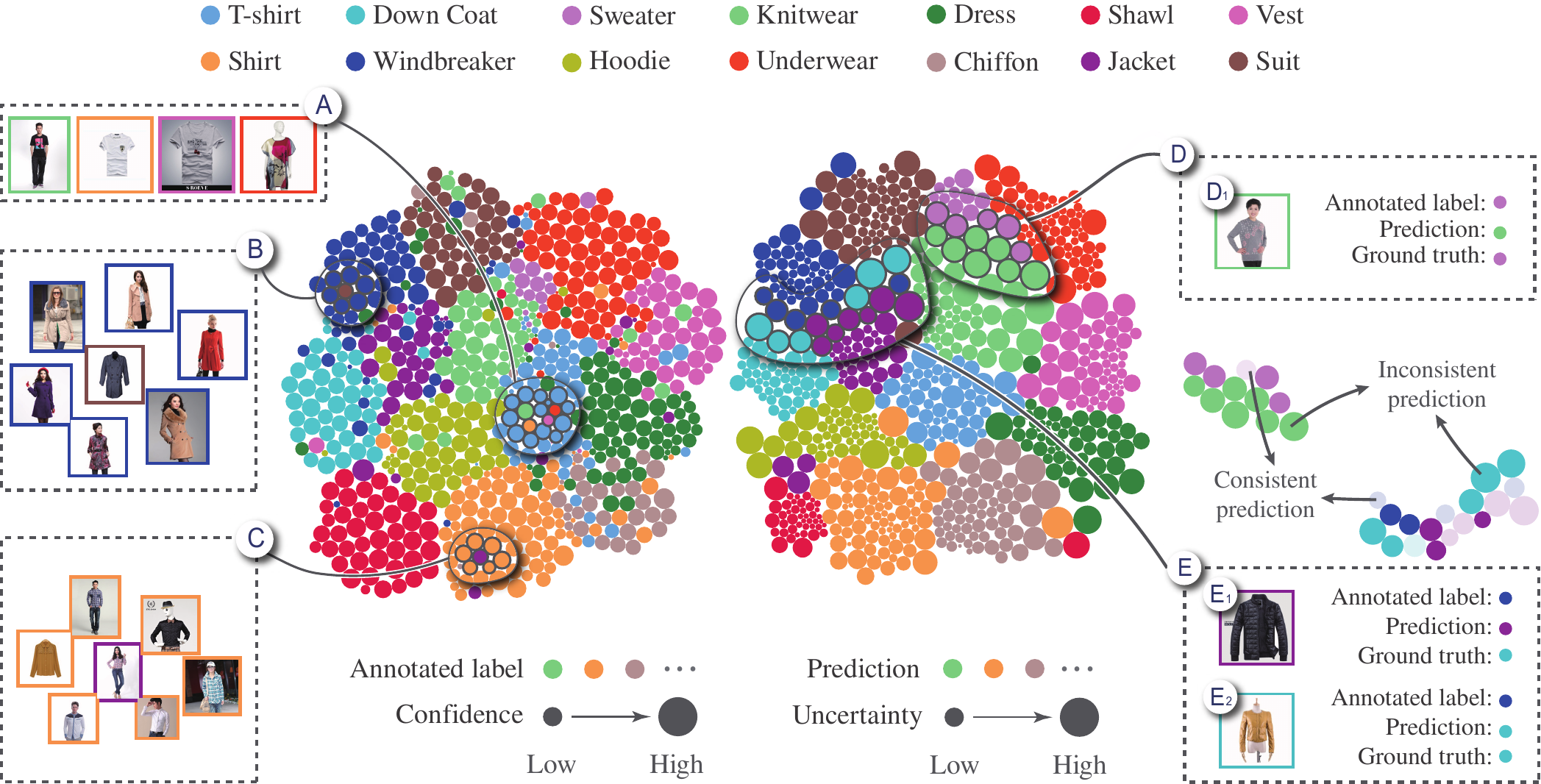}
  \put(-315, -10){(a)}
  \put(-185, -10){(b)}
\caption{Using neighborhood-preserving non-uniform circle packing to identify noisy labels in the sampled Clothing dataset: 
  (a) analyzing  images with random noise;
  (b) analyzing  images with content-ambiguity-related noise.
  }
  \label{fig:clothes}
\end{figure*}

Circle packing, which represents data items using tightly packed circles, has been widely used in many visualization applications due to its visually appealing and intuitive representation~\cite{abdullah2019malaysian,goertler2018bubble,wang2006circlepacking,wang2020visual,li2025routeflow,Liu2025}.
Recent research has shed light on the usefulness of neighborhood-preserving circle packing in data analysis.
For example, ArchExplorer~\cite{yuan2023archexplorer} represents neural network architectures as uniform circles of equal radii and places similar architectures adjacent to each other.
This helps users to compare the performance of architectures with similar structures and to gain insights into designing architectures.
However, ArchExplorer does not support non-uniform circles, which limits the analysis of quantitative attributes.
Consider the task of analyzing label noise in a clothing image dataset.
It is difficult to identify images mislabeled as similar classes, such as `knitwear' mislabeled as `sweaters', because they have consistent predictions with their neighboring images.
However, since these images usually have higher prediction uncertainty scores due to their ambiguous content, they can be easily identified by simultaneously encoding the uncertainty scores using radii.
For example,  larger circles along the cluster boundaries between the `knitwear' and `sweater' clusters (see \cref{fig:clothes}D) clearly highlight these ambiguous images.
This example highlights the need for a circle packing method that both preserves neighborhood relationships and encodes a quantitative attribute using radii.
In addition, previous studies concerning circle packing and cluster-aware methods highlight the importance of compactness of the layout and the convexity of clusters in data analysis~\cite{chen2021interactive, Rottmann2023MosaicSets, zhou2023cluster,zhou2025hierarchical}.

Based on the above analysis, we aim to simultaneously preserve neighborhood relationships and improve compactness and convexity, under the constraints that circles do not overlap and their radii encode the quantitative attribute of interest.
However, these objectives and constraints can conflict.
For example, improving compactness may compromise the ideal positions for neighborhood preservation.
As a result, the key challenge of neighborhood-preserving non-uniform circle packing lies in effectively balancing multiple optimization objectives while satisfying constraints.

To tackle this challenge, we formulate the neighborhood-preserving non-uniform circle packing problem as a maximal planar graph embedding problem.
In this graph, nodes represent circles, edges represent neighborhood relationships between circles, and the embedding provides the 2D coordinates of each node.
This formulation is based on the circle packing theorem~\cite{beardon1990uniformization}, which establishes a one-to-one correspondence between a non-uniform circle packing and a maximal planar graph.
Using this formulation, we develop a \emph{neighborhood-preserving non-uniform circle packing} method which we call NCP.
It is designed to effectively balance the objectives of neighborhood preservation, compactness, and convexity, while satisfying size and non-overlap constraints. 
To achieve this, we solve the associated multi-objective optimization problem with the continuation method~\cite{allgower2012numerical} that progressively incorporates different optimization objectives.
Initially, we project data items onto a 2D plane and generate a maximal planar graph and its initial embedding using the Delaunay triangulation.
This planar graph represents the maximal set of neighborhood relationships that we can preserve after  projection.
Then, we refine this embedding by progressively introducing additional optimization objectives for compactness and convexity while preserving neighborhood relationships.
This refinement integrates a power-diagram-based method for compactness and a force-directed method for convexity.
The final output is the planar graph embedding and the corresponding circle packing that simultaneously preserves neighborhood relationships and enhances compactness and convexity.
Our quantitative evaluation shows that \emph{NCP} can better preserve neighborhood relationships between data items than baseline methods.
Furthermore, \emph{NCP} achieves better convexity while obtaining comparable compactness.
The usability of our method is demonstrated in two use cases and a user study.
Our code is available at \url{https://github.com/NCP-2024/NCP}.
To sum up, the main contributions of this work are:
\begin{itemize}
    \item  formulation of neighborhood-preserving non-uniform circle packing as a maximal planar graph embedding problem,
    \item an optimization method that simultaneously preserves neighborhood relationships, and improves compactness and convexity, and
    \item an open-source library for generating neighborhood-preserving non-uniform circle packings.
\end{itemize}


\section{Related Work}
\label{sec:related-work}

Depending on whether spatial efficiency is optimized, circle-related layout methods can be divided into two groups: non-compact layout methods and compact layout (circle packing) methods.
Non-compact layout methods, such as the Dorling cartogram~\cite{sun2010effectiveness}, place the circles on a 2D plane without maximizing compactness.
In contrast, circle packing methods aim to optimize spatial efficiency by tightly packing circles, and are more closely related to our work.
As this is an NP-hard problem, many stochastic optimization methods have been developed~\cite{hifi2009review}, based on genetic algorithms~\cite{George1995Packing},  simulated annealing~\cite{zhang2004simulated}, and  adaptive beam search~\cite{akeb2009beam}.
However, these methods rely on a great deal of  trial-and-error, and require a considerable time to converge.
To speed up the process, faster heuristic methods have been proposed, which can be categorized into three classes: front-chain-based methods, power-diagram-based methods, and force-directed methods~\cite{scheibel2020survey}.

\emph{Front-chain-based methods} place each circle externally tangent to those already on the layout boundary~\cite{wang2006circlepacking}.
These boundary circles form the front chain of the layout.
G\"{o}rtler~\etal\cite{goertler2018bubble} utilized this method to generate a bubble treemap.
Researchers have also extended this method to visualize time series data, where preserving the temporal sequence of data items is crucial~\cite{liu2016online, zhao2014fluxflow}.
Front-chain-based methods incrementally place each circle without jointly considering their relative positions.
Thus, attempts to preserve neighborhoods may result in sub-optimal results.
This is demonstrated by our experiments in \cref{subsec:quantitative} (\emph{SimiFC}).
To tackle this issue,  power-diagram-based and force-directed methods simultaneously optimize the positions of all circles to preserve neighborhood relationships.

\emph{Power-diagram-based methods} employ a power diagram~\cite{aurenhammer1987power} to partition the layout region into non-overlapping cells and place circles within the corresponding cells~\cite{zhao2015variational}.
A power diagram is a weighted Voronoi diagram that partitions the layout region based on a set of weighted distances. 
By iteratively adjusting circle centers to agree with the maximum inscribed circles of their respective cells and increasing circle radii, the compactness of the resulting circle packing is improved.
Yu~\etal~\cite{yu2014content} utilized this method to create photo collages, where each photo was represented as a circle with a radius proportional to its importance.
A similar idea was also adopted by Liang~\etal~\cite{liang2018photo} and Rodrigues~\etal~\cite{rodrigues2023relaxed}.
\emph{Force-directed methods} treat circles as physical objects and employ simulated forces to pack them tightly.
Huron~\etal~\cite{samuel2013sedimentation} utilized gravitational forces and collision detection to generate a circle packing.
This method has been used in different applications such as enterprise analysis~\cite{liu2016social} and social media monitoring~\cite{Blumenstein2017Livevis}.

Both  power-diagram-based and force-directed methods have unique advantages in circle packing. 
Power-diagram-based methods achieve high compactness efficiently.
On the other hand,  force-directed methods provide a flexible framework for arranging circles based on simulated physical interactions, allowing the integration of various forces to meet different optimization objectives. 
Despite their advantages, each method also has limitations.
Power-diagram-based methods primarily focus on compactness and may not adequately preserve  neighborhood relationships between circles during the optimization process. 
However, without  good initialization, force-directed methods can face difficulties in balancing multiple competing objectives.
For example, attempting to optimize both compactness and neighborhood preservation can introduce forces with conflicting directions, which may lead to slower convergence and sub-optimal local minima.
This is demonstrated by our experiments in \cref{subsec:quantitative} (\emph{FD}). 
To address these issues, our method generates a neighborhood-preserving planar graph and then tries to preserve this planar graph during subsequent optimization. We first employ an improved power-diagram-based method that simultaneously considers compactness and neighborhood preservation.
The result is further refined using a force-directed method to obtain better convexity.
This hybrid method leverages the strengths of both  power-diagram and force-directed methods,  producing a better circle packing result.


\section{Optimization Objectives and Constraints}
\label{sec:guideline}

To generate a neighborhood-preserving non-uniform circle packing,
we distill the design criteria from existing circle packing and cluster-aware layout methods.
A detailed list is available in the supplemental materials.  
We organize these criteria on two levels: global and local.

\begin{figure}[t!]
    \begin{center}
    \includegraphics[width=\linewidth]{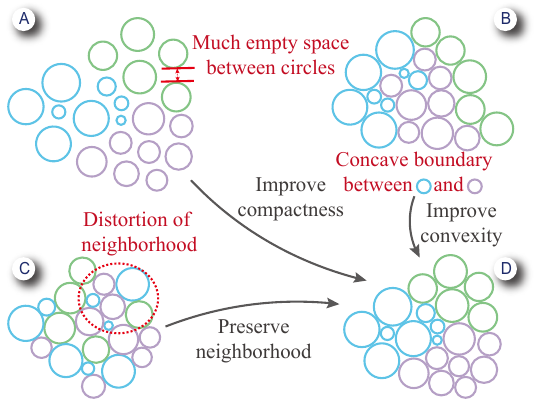}
    \caption{
        Optimization objectives for  neighborhood-preserving non-uniform circle packing.
    }
    \label{fig:measure}
    \end{center}
\end{figure}

The global criteria aim to generate a compact circle packing and clearly convey information from data.
Previous research has identified three criteria to achieve this goal~\cite{wang2006circlepacking, zhao2015variational}.
First, achieving high compactness (\emph{G1}, \cref{fig:measure}A) maximizes the utilization of the available display space~\cite{zhao2015variational}.
Second, circle radii typically encode quantitative attributes associated with data items (\emph{C1}), which enables  efficient comparison of  quantitative attributes~\cite{liu2016online, zhao2014fluxflow, yu2014content, meihoefer1969utility}.
Third, ensuring non-overlapping circles (\emph{C2}) reduces visual clutter, enhances the readability of individual circles, and simplifies data analysis~\cite{wang2006circlepacking, yuan2023archexplorer}.

The local criteria focus on two key aspects to facilitate data analysis: enhancing the perceptual clarity of clusters and preserving neighborhood relationships within each cluster.
Recent studies have demonstrated that optimizing the convexity of cluster shapes (\emph{G2}, \cref{fig:measure}B) can improve the perceptual clarity of clusters and thus facilitate  more efficient cluster analysis~\cite{chen2021interactive, zhou2023cluster}.
This also aligns with the Gestalt law of perceptual grouping~\cite{wagemans2012century}.
Additionally, preserving neighborhood relationships (\emph{G3}, \cref{fig:measure}C) enhances the user's ability to understand inherent structures in data and accurately identify outliers~\cite{xia2022interactive}.
\looseness=-1

Accordingly, our circle packing method employs three optimization objectives (\emph{G1}--\emph{G3}) and two hard constraints (\emph{C1}--\emph{C2}):
\begin{itemize}
\item\emph{G1: Compactness}. Create a tightly packed arrangement of circles to maximize spatial efficiency.

\item\emph{G2: Convexity}. Maintain a convex shape for each cluster to enhance perceptual clarity.

\item\emph{G3: Neighborhood preservation}.
Place circles that represent similar data items in close proximity within each cluster to preserve  neighborhood contexts.

\item\emph{C1: Size constraint}. Ensure that the radius of each circle is proportional to its quantitative attribute for comparative analysis. 

\item\emph{C2: Non-overlap constraint}. Arrange the circles without overlaps to maintain the distinctiveness of each circle.
\end{itemize}

\section{NCP Method}
\label{sec:formulation}

\begin{figure}[t]
    \begin{center}
    \includegraphics[width=\columnwidth]{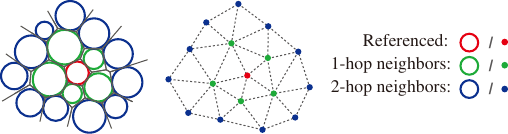}
    \put(-220, -10){(a)}
    \put(-135, -10){(b)}
    \caption{
        Correspondence between the circle packing result and the planar graph: (a)  example circle packing result, (b)  corresponding planar graph.
        Red circle: referenced circle, green circles: its $1$-hop neighbors, blue circles:  its $2$-hop neighbors.
    }   
    \label{fig:neighbor}
    \end{center}
\end{figure}

\begin{figure}[t]
    \begin{center}
    \includegraphics[width=\linewidth]{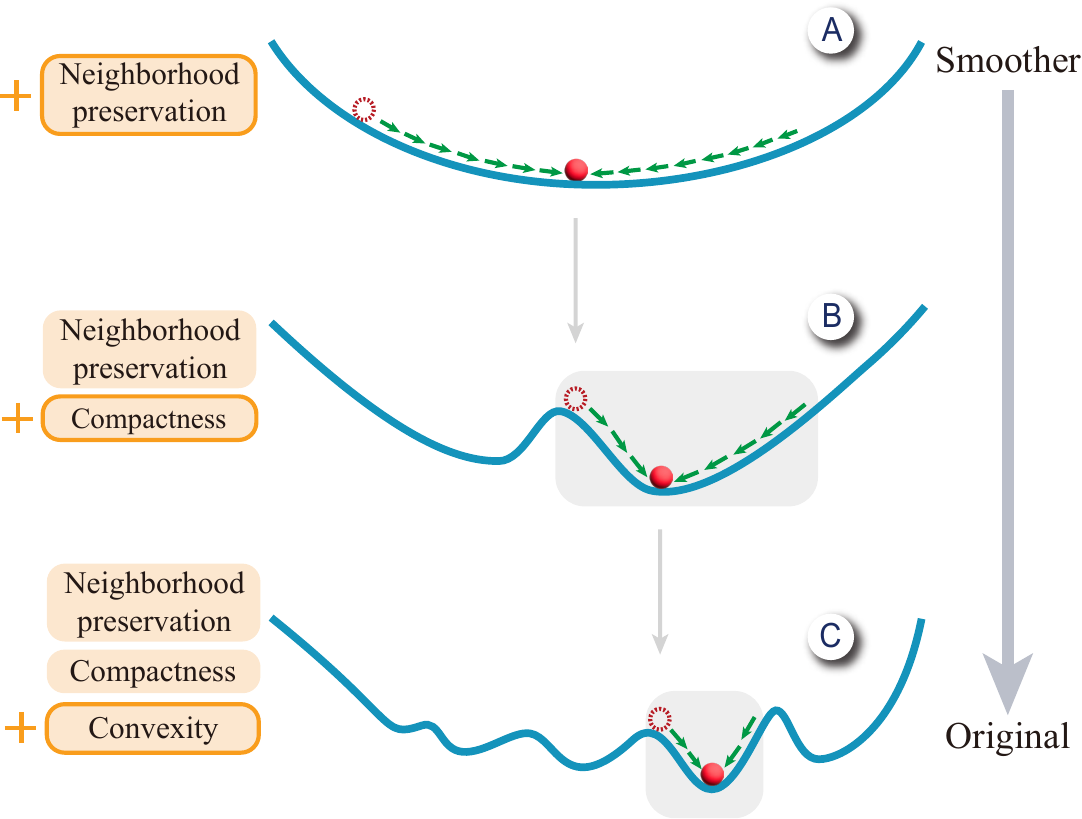}
    \caption{
        The continuation method's basic idea is to transform the original non-convex problem into a sequence of smoother ones and progressively solves them. This strategy guides optimization toward better regions in the solution space and leads to a better solution.
    }
    \label{fig:continuation}
    \end{center}
\end{figure}

\subsection{Planar-Graph-Based Problem Formulation}
According to the circle packing theorem~\cite{beardon1990uniformization}, a one-to-one correspondence exists between a non-uniform circle packing and a maximal planar graph.
In this graph, nodes represent circles, and edges represent neighborhood relationships between circles (\cref{fig:neighbor}).
Based on this correspondence, achieving an optimal circle packing is equivalent to identifying an appropriate planar graph and its embedding for placing circles.
Given $n$ data items and their associated quantitative attributes $\{w_i\}_{i=1}^n$ as input, 
our method generates a planar graph embedding and the corresponding circle packing that balances neighborhood preservation ($F_p$), compactness ($F_c$), and convexity ($F_v$), while satisfying the non-overlap ($C_o$) and size ($C_z$) constraints.
Following common practice, we adopt a weighted sum~\cite{marler2010weighted} to scalarize the multi-objective optimization problem as a single-objective problem:
\begin{equation}
\begin{aligned}
    &\argmin_{\{\bm{p}_i\}, s > 0} \quad  F_p + \alpha F_c + \beta F_v \\
    \text{such that} \quad 
    & r_i = s  w_i, \forall i, \ &(C_z), \\
    & r_i + r_j \le \|\bm{p}_i - \bm{p}_j\|, \forall i,  \  &(C_o),\\
    & 1 \le i \le n.
\label{eq:scalarized}
\end{aligned}
\end{equation}
Here, $\bm{p}_i$ is the center of the $i$-th circle, which is the embedding coordinate of the corresponding node in the planar graph.
The size constraint (C1) is inherently satisfied by optimizing the scaling factor $s$ that scales radii $r_i$ proportionally to $w_i$, while the non-overlap constraint (C2) can be achieved by reducing $s$ to prevent circle overlaps.
The parameters $\alpha$ and $\beta$, which balance the impact of the three terms, are determined by a grid search and set to $0.2$ and $1.0$ in our implementation.
In this setting, our method delivers near-optimal performance across all three goals.
A detailed sensitivity analysis is provided in the supplemental material.

\begin{figure*}[t!]
\centering
\includegraphics[width=\linewidth]{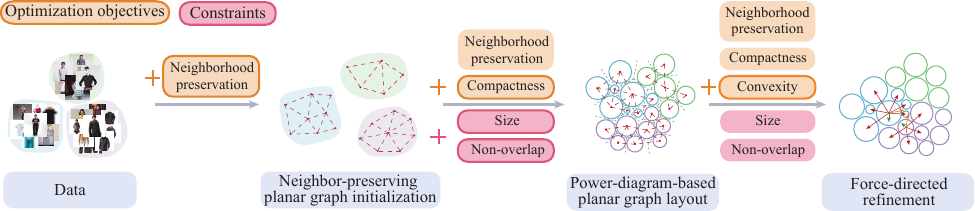}
\put(-337, -10){(a)}
\put(-180, -10){(b)}
\put(-45, -10){(c)}
\caption{Pipeline of the $\emph{NCP}$ method:
(a) \emph{neighborhood-preserving planar graph initialization} generates an initial planar graph that connects similar data items,
(b) \emph{power-diagram-based planar graph layout} refines the planar graph and produces an intermediate result preserving both compactness and neighborhood,
(c) \emph{force-directed refinement} compacts the circles and improves convexity.}
\label{fig:pipeline}
\end{figure*}

\subsection{Method Overview}
Despite having a well-defined optimization objective, the non-convex nature of the problem and the large number of local minima present a challenge~\cite{lu2011variational}.
An effective way to overcome this issue is the continuation method~\cite{allgower2012numerical}, which has been widely used in many learning tasks such as curriculum learning~\cite{bengio2009curriculum} and semi-supervised learning~\cite{chapelle2006continuation}.
As  \cref{fig:continuation} illustrates, the basic idea of this method is to transform the original non-convex problem into a sequence of smoother problems that are easier to solve.
Then, it sequentially solves them from the easiest one to the original one (\cref{fig:continuation}A--C); each solution serves as a starting point for solving the next problem~\cite{yang2010spectral}.
This strategy guides the optimization toward more favorable regions in the solution space and thus speeds convergence~\cite{bengio2009curriculum, wang2022survey, chen2024dynamic}.
Its effectiveness is demonstrated by a quantitative evaluation in \cref{subsec:quantitative}.

Accordingly, we design a three-step optimization method that progressively incorporates our three objectives.
The pipeline is shown in \cref{fig:pipeline}.
In the first step, \emph{neighborhood-preserving planar graph initialization} (\cref{fig:pipeline}(a)), 
we create a maximal planar graph that preserves neighborhood relationships as well as possible by maximizing connections between similar data items. 
This initial graph serves as the starting point for  subsequent optimization.
The  two subsequent steps involve refining the embedding of the graph to optimize compactness and convexity while preserving neighborhood relationships. 
Compactness is prioritized over convexity because  it is affected by all circles, whereas convexity is only affected by circles on cluster boundaries~\cite{zhou2023cluster}. 
Thus, in the second step, \emph{power-diagram-based planar graph layout} (\cref{fig:pipeline}(b)), we generate an intermediate result to balance compactness and neighborhood preservation.
In the third step, \emph{force-directed refinement} (\cref{fig:pipeline}(c)), we improve convexity while also preserving neighborhood relationships and compactness by using simulated forces to adjust the circle positions.

\subsection{Neighborhood-Preserving Planar Graph Initialization}
\label{subsec:planar-init}
To generate a neighborhood-preserving planar graph, we begin by using a projection method to place similar data items in close proximity.
We then generate the edges using Delaunay triangulation, which is effective in connecting similar data items without intersections~\cite{misue1995layout}.
In this way, we generate a maximal planar graph in which  edges represent  neighborhood relationships between data items, to be preserved during the following steps.

The projection method is crucial because it decides which neighborhood relationships  will be preserved.
To select the most suitable method,
we  conducted an experiment on eight high-dimensional datasets with cluster structures, as used in Xia~\etal's work~\cite{xia2021revisiting}.
We identified five candidate projection methods: t-SNE~\cite{van2008visualizing}, UMAP~\cite{mcinnes2018umap}, PCA~\cite{wold1987principal}, MDS~\cite{kruskal1978multidimensional}, and NMF~\cite{lee1999learning}, based on previous studies~\cite{vernier2020quantitative, xia2021revisiting}. 
To evaluate the ability of these projection methods to preserve neighborhoods, we used the neighborhood preservation degree~\cite{kruiger2017graph, zhong2023force}, which calculates how many neighborhoods in the high-dimensional space are preserved as neighbors in the generated planar graph.
\begin{equation}
F_p = -\frac{1}{n}\sum_{i=1}^{n} \frac{|\Gamma_G(i) \cap \Gamma_D(i, k_i)|}{|\Gamma_G(i) \cup \Gamma_D(i, k_i)|}, \quad k_i=|\Gamma_G(i)|,
\label{eq:np}
\end{equation}
where $G$ is the generated planar graph, and $\Gamma_G(i) = \{j \mid d_G(i, j) = 1, l_i = l_j, \forall 1 \le j \le m, j \neq i\}$ denotes the $1$-hop neighbors of the $i$-th data item with the same cluster label.
$d_G(i, j)$ is the graph distance between the $i$-th and $j$-th data items, and $l_i$ and $l_j$ are their cluster labels.
$\Gamma_D(i, k_i)$ denotes the $k_i$ nearest neighbors of the $i$-th data item in the high-dimensional space based on data similarity.

The experimental results show that t-SNE best preserves  neighborhoods, achieving an average neighborhood preservation degree of 0.392.
Notably, t-SNE yielded the best result in seven out of eight cases, and ranked second in the remaining case.
The circle packing results for various initialization methods are also available in the supplemental material, further illustrating t-SNE’s qualitative advantages.
Therefore, we use t-SNE in our implementation.
Users could instead employ other methods based on their analysis needs.

\subsection{Power-Diagram-Based Planar Graph Layout}
\label{subsec:power-diagram}

Following the optimization of neighborhood preservation, the second step aims to improve compactness with the size and non-overlap constraints.
To meet this additional objective and the constraints, we refine the embeddings of the previously generated neighborhood-preserving planar graph:
\begin{equation}
\begin{aligned}
\argmin_{\{\bm{p}_i\}_{i=1}^n, s > 0} \quad &  F_p + \alpha F_c \\
\text{such that} \quad & C_z, C_o,
\end{aligned}
\label{eq:power-diagram}
\end{equation}
The first term ($F_p$) denotes neighborhood preservation, and the second term ($F_c$) denotes compactness.
For the first term, instead of directly optimizing the discrete term $F_p$ as defined in \cref{eq:np}, we use a continuous form $F_p^{'}$ to facilitate optimization.
Here, $F_p^{'} = \sum_{i=1}^m \sum_{j \in \Gamma_G(i)} {\|\bm{p}_i -\bm{p}_j\|}/{s}$, which
encourages  circle pairs connected by  edges to remain close.
For the second term, we follow the method in MosaicSets~\cite{Rottmann2023MosaicSets} and use $F_c = \sum_{i=1}^m {\|\bm{p}_i-\bm{O}\|}/{s}$, which encourages  circles to be placed near the center $\bm{O}$ to enhance compactness.

\begin{figure}[t]
    \centering
    \includegraphics[width=\columnwidth]{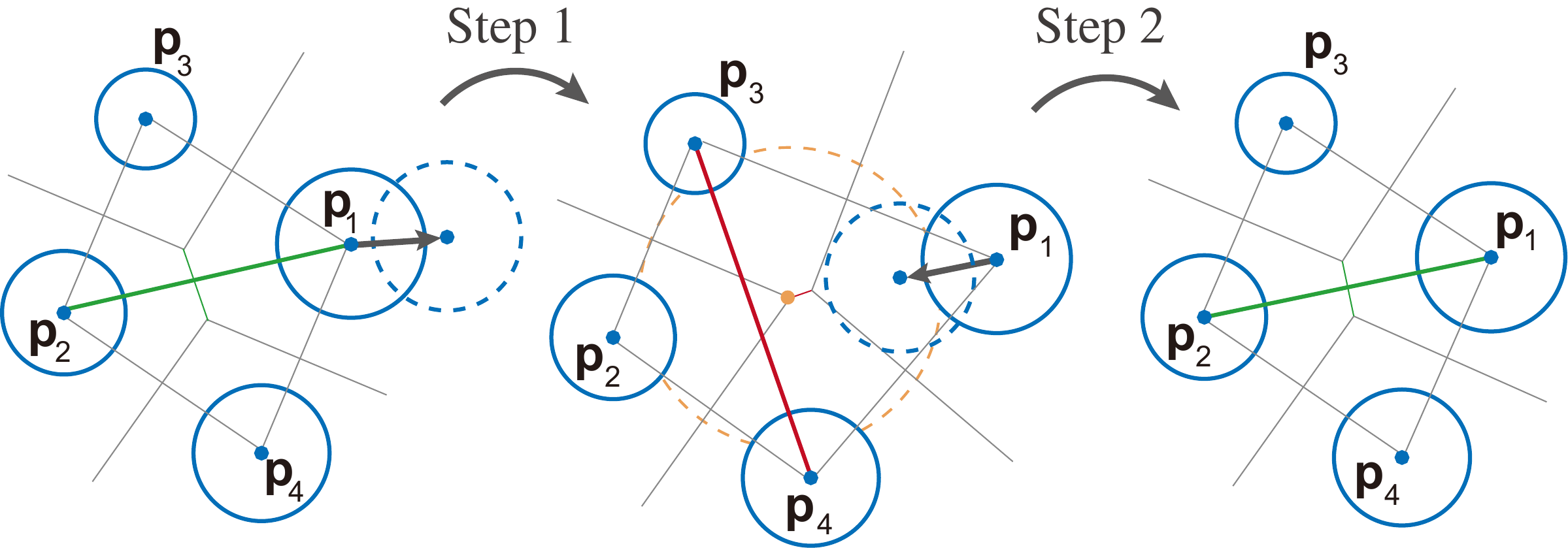}
    \put(-220, -10){(a)}
    \put(-130, -10){(b)}
    \put(-42, -10){(c)}
    \caption{Enhancing  neighborhood preservation in the power-diagram-based algorithm: (a) initial edge $\bm{p}_1\bm{p}_2$ (green); (b)  node movements in Step 1 cause an edge change: the original edge $\bm{p}_1\bm{p}_2$ (green) is removed, and a new edge $\bm{p}_3\bm{p}_4$ (red) is added; (c) after moving $\bm{p}_1$ in the Step 2, the change in edges is mitigated.}
    \label{fig:add-move}
\end{figure}

To solve the optimization problem defined in \cref{eq:power-diagram}, we leverage the power-diagram-based method developed by Zhao~\etal~\cite{zhao2015variational} to improve compactness while ensuring the constraints.
This method moves  nodes toward the centers of the maximum inscribed circles within their respective Voronoi cells and enlarges the scaling factor.
However, this may change neighborhood relationships between circles, which leads to the removal or addition of  corresponding edges in the planar graph.
Fig.~\ref{fig:add-move} provides an illustrative example where the original edge $\bm{p}_1 \bm{p}_2$ (in green) is removed, and a new edge $\bm{p}_3 \bm{p}_4$ (in red) is added after moving $\bm{p}_1$.
This is because $\bm{p}_1$ now lies outside the weighted circumcircle of $\bm{p}_2 \bm{p}_3 \bm{p}_4$ (the orange dotted circle)~\cite{Goes2014Weighted}.
To address this issue, we move $\bm{p}_1$ towards the center of their corresponding weighted circumcircles $\bm{p}_2 \bm{p}_3 \bm{p}_4$ (Figs.~\ref{fig:add-move}(b)--(c)).
For each adjustment, we start from a predefined movement distance and then iteratively reduce it by half until the objective function ($F_p^{'} + \alpha F_c$) is improved.
This ensures  fast convergence when alternatively optimizing compactness and neighborhood preservation.

\begin{figure}[t]
    \centering
    \includegraphics[width=\columnwidth]{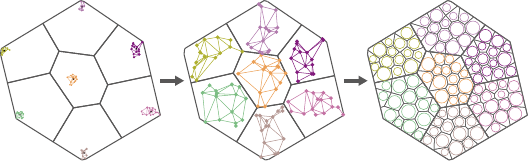}
    \put(-216, -10){(a)}
    \put(-129, -10){(b)}
    \put(-42, -10){(c)}
    \caption{Parallel computation of power diagrams: (a) cluster nodes based on their initial positions and partition the layout region; (b) redistribute the nodes in each cluster with 2D transformations; (c) construct the power diagram for each cluster.}
    \label{fig:div-conq}
\end{figure}

In addition, we propose two practical improvements to accelerate this method.
First, given the time-consuming nature of calculating the maximum inscribed circle center within a cell, we approximate it by calculating its centroid, which is less computationally demanding~\cite{Passenbrunner2011SkyObserver}. 
Second, since \emph{NCP} focuses on preserving neighborhood relationships within each cluster, we parallelize the computation of power diagrams across different clusters (\cref{fig:div-conq}).
We start by clustering the nodes based on their positions using $k$-means~\cite{lloyd1982least}, with the optimal number of clusters selected via the silhouette score~\cite{rousseeuw1987silhouettes}.
The layout region is divided into several sub-regions based on the clustering result (\cref{fig:div-conq}(a)).
To improve compactness, we restrict the size of each sub-region proportionally to the total area of circles within each cluster using the capacity-constraint power diagram~\cite{aurenhammer1998minkowski}.
To more evenly distribute the corresponding nodes in each sub-region, we apply 2D transformations, including translation, rotation, and scaling, to the sub-graph within each cluster (\cref{fig:div-conq}(b)), aiming to enlarge it and fill the sub-region as much as possible.
These transformations preserve the edges in the initial planar graph and minimize the need for additional node movements in subsequent optimization, which contributes to better neighborhood preservation.
Finally, the power diagram of each cluster is generated concurrently (\cref{fig:div-conq}(c)).

\subsection{Force-Directed Refinement}
\label{subsec:force-directed}
The final step incorporates convexity, which brings us to the original problem of jointly optimizing all objectives while satisfying the size and non-overlap constraints:
\begin{equation}
\begin{aligned}
    \argmin_{\{\bm{p}_i\}_{i=1}^n} \quad & F_p^{'}+\alpha F_c+\beta F_v, \\
    \text{such that} \quad & C_z, C_o.
\end{aligned}
\label{eq:compaction}
\end{equation}

\begin{figure}[t!]
\centering
\includegraphics[width=\linewidth]{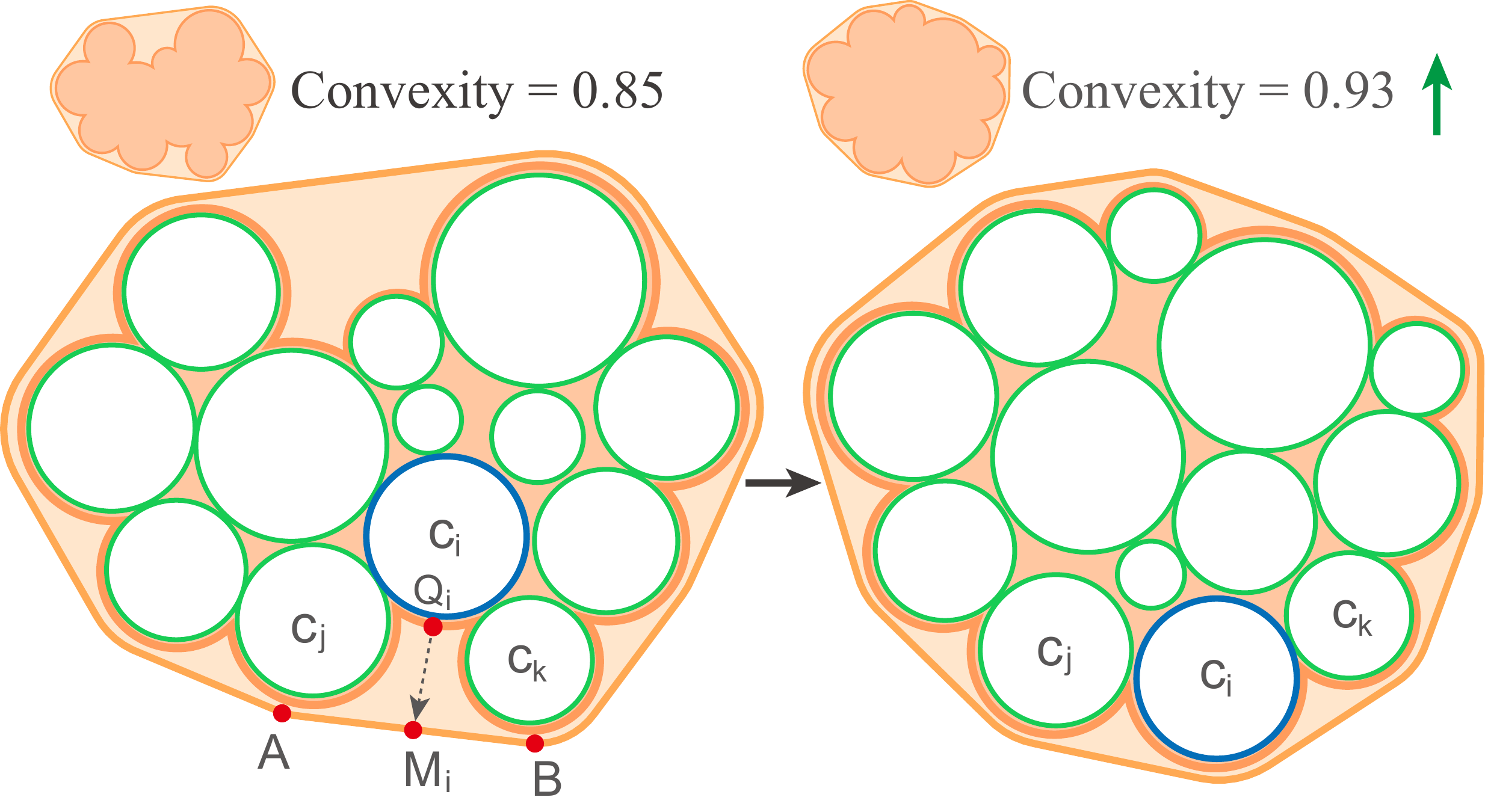}
\put(-163, -9){(a)}
\put(-52, -9){(b)}
\caption{Example of improving the convexity of cluster shapes: (a) move the blue circle on the cluster boundary toward the convex hull of the cluster; (b) calculate the movement direction of the blue circle to form a more convex boundary.}
\label{fig:convexity-force}
\end{figure}

The first two terms encourage neighborhood preservation ($F_p^{'}$) and compactness ($F_c$), as defined in the second step.
The third term ($F_v$) is introduced to promote the formation of convex cluster shapes by adjusting the circles on cluster boundaries.
Inspired by EulerSmooth~\cite{Simonetto2016Simple}, the basic idea is to move the circles on the cluster boundary toward the convex hull, such as the blue circle $c_i$ in \cref{fig:convexity-force}(a). 
The key is to determine the movement directions for these circles.
For each circle on the cluster boundary, we identify two neighboring circles on the left and right that are also on the boundary and intersect  the convex hull.
The goal is to improve the convexity in this local region without compromising  neighborhood preservation.
To achieve this, we move the circle to close this gap while minimizing  movements of other circles to preserve the neighborhood structure.
For example, in \cref{fig:convexity-force}(a), we move $c_i$ to close the gap between adjacent circles $c_j$ and $c_k$ that touch  the convex hull. 
These two circles require only minimal movement to accommodate $c_i$ between them.
Specifically, suppose the convex hull touches  $c_j$ and $c_k$ at points $A$ and $B$, with $M_i$ representing the midpoint of $AB$.
We identify the point $Q_i$ on $c_i$ that is closest to $M_i$.
We then move $c_i$ from $Q_i$ towards $M_i$.
Accordingly, $F_v = \sum_{i \in \delta} \|Q_i - M_i\|$, where $\delta$ represents the set of circles on the cluster boundaries (\cref{fig:convexity-force}(b)).

To solve the optimization problem defined in \cref{eq:compaction}, we employ a force-directed method to refine the circle positions.
Here, the forces are set to the negative gradient of the objective function.
To avoid overlaps between circles, we utilize a 2D physics engine, Box2D~\cite{catto2010box2d}, to simulate forces.
The simulation velocity is adjusted using a linear annealing decay for better convergence.

\section{Evaluation}
\label{sec:evaluation}
In this section, we first present a quantitative evaluation to show the effectiveness of the proposed \emph{NCP} method.
We then present two use cases to demonstrate how \emph{NCP} can be used to facilitate data analysis.

\subsection{Quantitative Evaluation}
\label{subsec:quantitative}

\subsubsection{Experimental Setting}
\label{subsubsec:exp_settings}

\paragraph{Datasets}
The quantitative evaluation was conducted on eight high-dimensional datasets with cluster structures, previously used in Xia~\etal's work~\cite{xia2021revisiting}.
These datasets vary in size from 155 to 1083 data points with 7 to 192  dimensions.
In practical applications, the number of samples simultaneously displayed in non-uniform circle packing typically does not exceed one thousand~\cite{abdullah2019malaysian}.
Displaying more samples reduces the circle radii, making it difficult to discern  radius differences on standard monitors.
Similarity between  data items was computed based on the features provided in the datasets.
The radii of the circles were generated from two distinct uniform distributions, one spanning from $0.1$ to $1$ and the other from $0.5$ to $1$.
The two distributions were employed to simulate  cases where the circle radii had  large or small variance.

\begin{table*}[t]
    \centering
    \caption{Performance comparison in terms of neighborhood preservation ($NP_1$, $NP_2$), compactness, and convexity.
    }\label{tab:performance}
    
   \begin{subtable}[t]{\linewidth}
   \captionsetup{font={normal}}
   \setlength\tabcolsep{3pt}
    \begin{tabular}{l|ccccccccccccccccccccc}
       \toprule
        \multirow{3}{*}{\textbf{Dataset}}
        & \multicolumn{4}{c}{${NP_1}$}
        & \multicolumn{4}{c}{${NP_2}$}
        & \multicolumn{4}{c}{{Compactness}}
        & \multicolumn{4}{c}{{Convexity}} \\
        \cmidrule(lr){2-5} \cmidrule(lr){6-9} \cmidrule(lr){10-13} \cmidrule(lr){14-17}
        & \multicolumn{2}{c}{Baseline} & \multicolumn{2}{c}{Ours}
        & \multicolumn{2}{c}{Baseline} & \multicolumn{2}{c}{Ours}
        & \multicolumn{2}{c}{Baseline} & \multicolumn{2}{c}{Ours}
        & \multicolumn{2}{c}{Baseline} & \multicolumn{2}{c}{Ours} \\
        \cmidrule(lr){2-3} \cmidrule(lr){4-5} \cmidrule(lr){6-7} \cmidrule(lr){8-9} \cmidrule(lr){10-11} \cmidrule(lr){12-13} \cmidrule(lr){14-15} \cmidrule(lr){16-17}
        & {SimiFC} & {AEF} & {FD} & {NCP}
        & {SimiFC} & {AEF} & {FD} & {NCP}
        & {SimiFC} & {AEF} & {FD} & {NCP}
        & {SimiFC} & {AEF} & {FD} & {NCP} \\
        \midrule
        Boston &
        0.255 & 0.235 & 0.285 & \textbf{0.341} &
        0.270 & 0.287 & 0.394 & \textbf{0.423} &
        0.865 & 0.882 & \textbf{0.889} & \textbf{0.889} &
        0.507 & 0.665 & \textbf{0.837} & 0.809 \\
        
        Dermatology & 
        0.220 & 0.256 & 0.290 & \textbf{0.311} &
        0.238 & 0.278 & 0.370 & \textbf{0.385} &
        0.858 & 0.868 & \textbf{0.876} & 0.874 &
        0.582 & 0.683 & \textbf{0.796} & 0.790 \\
        
        Ecoli & 
        0.248 & 0.233 & 0.298 & \textbf{0.344} &
        0.242 & 0.271 & 0.378 & \textbf{0.411} &
        0.845 & 0.876 & \textbf{0.880} & 0.874 &
        0.633 & 0.821 & 0.802 & \textbf{0.831} \\ 

        ExtYaleB &
        0.371 & 0.321 & 0.406 & \textbf{0.422} &
        0.364 & 0.355 & 0.476 & \textbf{0.487} &
        0.850 & \textbf{0.876} & 0.872 & 0.868 &
        0.590 & 0.782 & 0.788 & \textbf{0.806} \\ 
        
        MNIST64 &
        0.203 & 0.203 & 0.276 & \textbf{0.306} &
        0.207 & 0.224 & 0.349 & \textbf{0.372} &
        0.833 & 0.866 & \textbf{0.867} & 0.863 &
        0.455 & 0.713 & 0.821 & \textbf{0.827} \\
        
        Olive &
        0.210 & 0.199 & 0.271 & \textbf{0.328} &
        0.213 & 0.235 & 0.382 & \textbf{0.414} &
        0.844 & 0.866 & \textbf{0.875} & 0.862 &
        0.554 & 0.713 & 0.784 & \textbf{0.817} \\
        
        Weather &
        0.369 & 0.320 & 0.352 & \textbf{0.448} &
        0.355 & 0.366 & 0.527 & \textbf{0.567} &
        0.850 &0.872 & \textbf{0.877} & 0.868 &
        0.389 & 0.683 & \textbf{0.843} & 0.795 \\
        
        World12D &
        0.194 & 0.253 & 0.368 & \textbf{0.379} &
        0.252 & 0.305 & 0.462 & \textbf{0.463} &
        0.866 & 0.888 & \textbf{0.900} & {0.896} &
        0.596 & 0.630 &0.782 & \textbf{0.809} \\
        
        \midrule
        
        \textbf{Average} &
        0.259 & 0.253 & 0.318 & \textbf{0.360} & 
        0.268 & 0.290 & 0.417 & \textbf{0.440} & 
        0.851 & 0.874 & \textbf{0.880} & 0.874 & 
        0.538 & 0.711 & 0.807 & \textbf{0.811} \\ 
        
        \bottomrule
    \end{tabular}
    \caption{Performance when circle radii have a large variance.}
    \end{subtable} 

    \begin{subtable}[t]{\linewidth}
    \captionsetup{font={normal}}
    \setlength\tabcolsep{3pt}
    \begin{tabular}{c|ccccccccccccccccccccc}
       \toprule
        \multirow{3}{*}{\textbf{Dataset}}
        & \multicolumn{4}{c}{${NP_1}$}
        & \multicolumn{4}{c}{${NP_2}$}
        & \multicolumn{4}{c}{{Compactness}}
        & \multicolumn{4}{c}{{Convexity}} \\
        \cmidrule(lr){2-5} \cmidrule(lr){6-9} \cmidrule(lr){10-13} \cmidrule(lr){14-17}
        & \multicolumn{2}{c}{Baseline} & \multicolumn{2}{c}{Ours}
        & \multicolumn{2}{c}{Baseline} & \multicolumn{2}{c}{Ours}
        & \multicolumn{2}{c}{Baseline} & \multicolumn{2}{c}{Ours}
        & \multicolumn{2}{c}{Baseline} & \multicolumn{2}{c}{Ours} \\
        \cmidrule(lr){2-3} \cmidrule(lr){4-5} \cmidrule(lr){6-7} \cmidrule(lr){8-9} \cmidrule(lr){10-11} \cmidrule(lr){12-13} \cmidrule(lr){14-15} \cmidrule(lr){16-17}
        & {SimiFC} & {AEF} & {FD} & {NCP}
        & {SimiFC} & {AEF} & {FD} & {NCP}
        & {SimiFC} & {AEF} & {FD} & {NCP}
        & {SimiFC} & {AEF} & {FD} & {NCP} \\
        \midrule
        Boston &
        0.257 & 0.232 & 0.315 & \textbf{0.365} &
        0.251 & 0.260 & 0.396 & \textbf{0.437} & 
        0.851 & {0.870} & 0.881 & \textbf{0.888} &
        0.557 & 0.657 & \textbf{0.842 }& 0.833 \\
        
        Dermatology & 
        0.233 & 0.284 & 0.319 & \textbf{0.364} &
        0.259 & 0.278 & 0.378 & \textbf{0.399} &
        0.845 & {0.868} & 0.872 & \textbf{0.875} &
        0.589 & 0.691 & 0.842 & \textbf{0.845} \\
        
        Ecoli & 
        0.222 & 0.257 & \textbf{0.360} & 0.350 &
        0.236 & 0.286 & 0.415 & \textbf{0.416} & 
        0.841 & {0.862} & \textbf{0.870} & 0.866 &
        0.516 & 0.767 & 0.820 & \textbf{0.821} \\
        
        ExtYaleB &
        0.393 & 0.375 & 0.442 &\textbf{ 0.476} &
        0.349 & 0.392 & 0.486 & \textbf{0.505} &
        0.842 & 0.864 & \textbf{0.870} & 0.862 & 
        0.500 & 0.757 & 0.801 & \textbf{0.808} \\   
        
        MNIST64 &
        0.215 & 0.235 & 0.274 & \textbf{0.327} &
        0.203 & 0.231 & 0.335 & \textbf{0.375} & 
        0.828 & 0.856 & \textbf{0.874} & 0.854 & 
        0.460 & 0.720 & \textbf{0.828} & 0.808 \\ 
        
        Olive &
        0.219 & 0.235 & 0.243 & \textbf{0.342} &
        0.215 & 0.245 & 0.322 & \textbf{0.422} &
        0.838 & {0.861} & \textbf{0.870} & 0.858 & 
        0.489 & 0.615 & 0.809 & \textbf{0.839} \\
        
        Weather &
        0.408 & 0.355 & 0.177 & \textbf{0.483} &
        0.363 & 0.365 & 0.232 & \textbf{0.578} &
        0.837 & 0.865 & \textbf{0.891} & 0.860 &
        0.366 & 0.658 & 0.743 & \textbf{0.816} \\   
        
        World12D &
        0.306 & 0.305 & \textbf{0.399} & 0.380 &
        0.310 & 0.320 & 0.476 & \textbf{0.477} &
        0.857 & {0.870} & \textbf{0.896} & 0.885 &
        0.480 & 0.632 & \textbf{0.800} & 0.796 \\
        
        \midrule
        
        \textbf{Average} &
        0.281 & 0.285 & 0.316 & \textbf{0.386} & 
        0.273 & 0.297 & 0.380 & \textbf{0.451} & 
        0.842 & 0.864 & \textbf{0.878} & 0.868 & 
        0.495 & 0.687 & 0.810 & \textbf{0.825} \\ 
        
        \bottomrule
    \end{tabular}
    \caption{Performance when circle radii have a small variance.}
    \end{subtable} 

\end{table*}

\paragraph{Baseline methods}
As no existing circle packing methods simultaneously consider non-uniform circles and neighborhood preservation, 
we extended two representative circle packing methods to provide baselines.
In addition, we considered a force-directed method, to demonstrate the effectiveness of the continuation method.
We did not include a power-diagram-based method because it fails to produce a compact circle packing result.

\emph{SimiFC} ({Simi}larity-aware {F}ront-{C}hain). 
We extended the front-chain-based method~\cite{wang2006circlepacking} to improve neighborhood preservation by improving its strategy for placing circles.
In each iteration, it first identifies the position outside the front chain that is closest to the layout center and then places the circle that maximizes neighborhood preservation.

\emph{AEF} ({A}rch{E}xplorer + {F}orce-directed).
The neighborhood-preserving circle packing method in ArchExplorer places uniform circles on a hexagonal grid and then greedily swaps them to maximize neighborhood preservation.
We extended this method to support the packing of non-uniform circles by combining it with the force-directed method. 
The extension is referred to as \emph{AEF}.
\emph{AEF} initially determines the positions of circles using the original method without considering their radii.
The positions are then refined using the force-directed method described in \cref{subsec:force-directed}.
\looseness=-1

\emph{FD} ({F}orce-{D}irected).
This basic method  skips the power-diagram-based planar graph layout step.
Specifically, we first placed  circles based on  t-SNE projection and then directly optimized their positions using the force-directed method.
The weights in the optimization objective \cref{eq:compaction} are the same as those used in the \emph{NCP} method.\looseness=-1

\paragraph{Parameters}
The perplexity of t-SNE was set to $15$ to give the best neighborhood preservation, as determined by a grid search.
The force-directed method ran $1,250$ iterations for \emph{NCP} and \emph{AEF}, and $10,000$ iterations for \emph{FD} to guarantee convergence on all datasets.

\begin{figure}[t]
    \begin{center}
    \includegraphics[width = \linewidth]{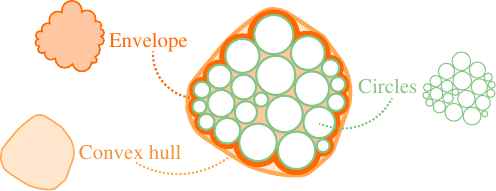}
   \caption{
        Envelope (thick orange curve) and  convex hull (thin orange curve) of a set of tightly packed circles.
    }
    \label{fig:hull}
    \end{center}
\end{figure}

\subsubsection{Measures}
We evaluated the circle packing results based on neighborhood preservation, compactness, and convexity.
The scores of these measures range from $0$ to $1$; higher scores indicate better results.

\paragraph{Neighborhood preservation}
This measure quantifies  neighbor overlap between the high-dimensional data and their 2D embeddings.
It is defined based on the neighborhood preservation degree from \cref{eq:np}:
$$
NP_1 = \frac{1}{m}\sum_{i=1}^{m} \frac{|\Gamma_G(i) \cap \Gamma_D(i, k_i)|}{|\Gamma_G(i) \cup \Gamma_D(i, k_i)|}, \quad k_i=|\Gamma_G(i)|.
$$
Moreover, to assess the capability to preserve a larger range of neighborhoods, we introduce $NP_2$ to consider  $2$-hop neighbors of each node in the planar graph:
$$
NP_2 = \frac{1}{m}\sum_{i=1}^{m} \frac{|\Gamma_G^{'}(i) \cap \Gamma_D(i, k_i^{'})|}{|\Gamma_G^{'}(i) \cup \Gamma_D(i, k_i^{'})|}, \quad k_i^{'}=|\Gamma_G^{'}(i)|.
$$
Here $\Gamma_G^{'}(i)=\{j \mid d_G(i, j) \le 2, l_i = l_j, \forall 1 \le j \le m, j \neq i \}$ is the set of $2$-hop neighbors of the $i$-th node.

\paragraph{Compactness}
We evaluated compactness following Liang~\etal~\cite{liang2018photo}.
Compactness is defined as the ratio of the total area covered by all circles to the area of their envelope:
$$
\mathrm{Compactness} = \frac{\Area(\cup_i c_i)}{\Area(\Enve(\cup_i c_i))},
$$
where $c_i$ denotes the $i$-th circle.
For any region $\Omega$, $\Area(\Omega)$ denotes its area.
$\Enve(\Omega)$ denotes its envelope, which is defined as the shape formed by its outer boundary~\cite{goertler2018bubble} (see the thick orange curve in \cref{fig:hull}).

\paragraph{Convexity}
We evaluated convexity based on a popular measure that is defined as the ratio of the area of a shape's envelope to that of its convex hull~\cite{zunic2004new}.
Since circles within a cluster may be divided into multiple connected components in the results, the envelope of a cluster is defined as the union of the envelopes of such connected components.
The convex hull of a cluster is defined in the same manner.
Consequently, the convexity of the $j$-th cluster is expressed as $\Area(\Enve(\cup_{l_i=j} c_i))\ /\ \Area(\Convexhull(\cup_{l_i=j} c_i))$, 
where $\Convexhull(\Omega)$ denotes the convex hull of $\Omega$  (the light orange curve in \cref{fig:hull}).
The convexity of a circle packing is defined as the average convexity across all clusters:
$$
\mathrm{Convexity} = \frac{1}{L} \sum_{j=1}^L \frac{\Area(\Enve(\cup_{l_i=j} c_i))}{\Area(\Convexhull(\cup_{l_i=j} c_i))},
$$
where $L$ is the number of clusters.

\subsubsection{Results}
Our experimental results are presented in \cref{tab:performance}.
Overall, our method (\emph{NCP}) achieved performed better  than the two baseline methods (\emph{SimiFC} and \emph{AEF}) in terms of neighborhood preservation and convexity while showing comparable performance on compactness.
The reason for the lower neighborhood preservation and convexity of the two baseline methods is that they adopt greedy strategies  instead of  projection methods to place circles.
Such strategies result in suboptimal circle arrangements for preserving neighborhood relationships and do not guarantee the formation of convex cluster shapes.
Consequently, we mainly focus on the comparison between \emph{NCP} and the basic method \emph{FD}.

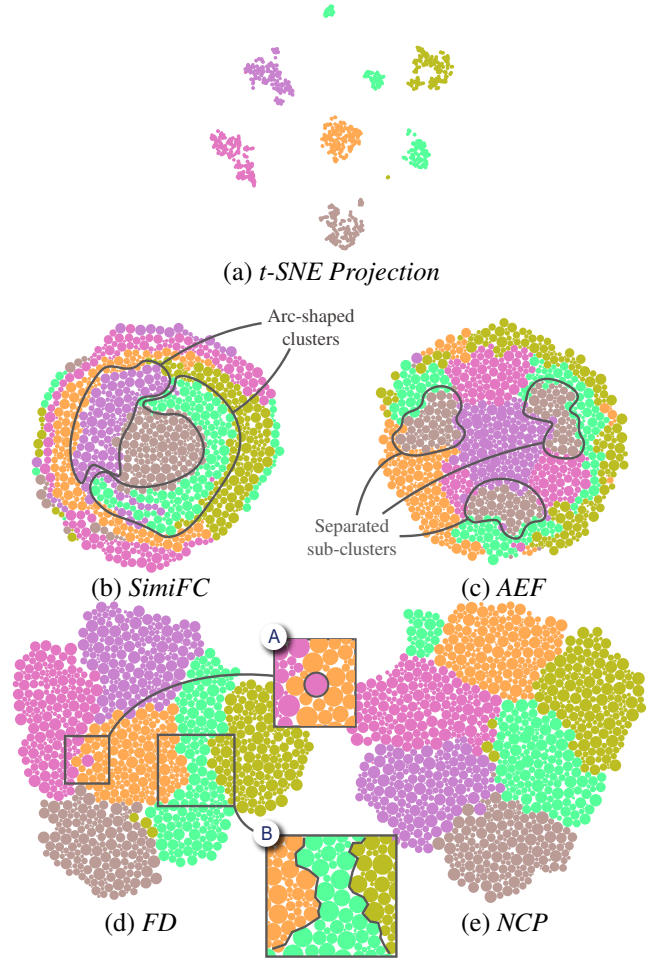
\begin{figure}[t]
\centering
\includegraphics[width=\linewidth]{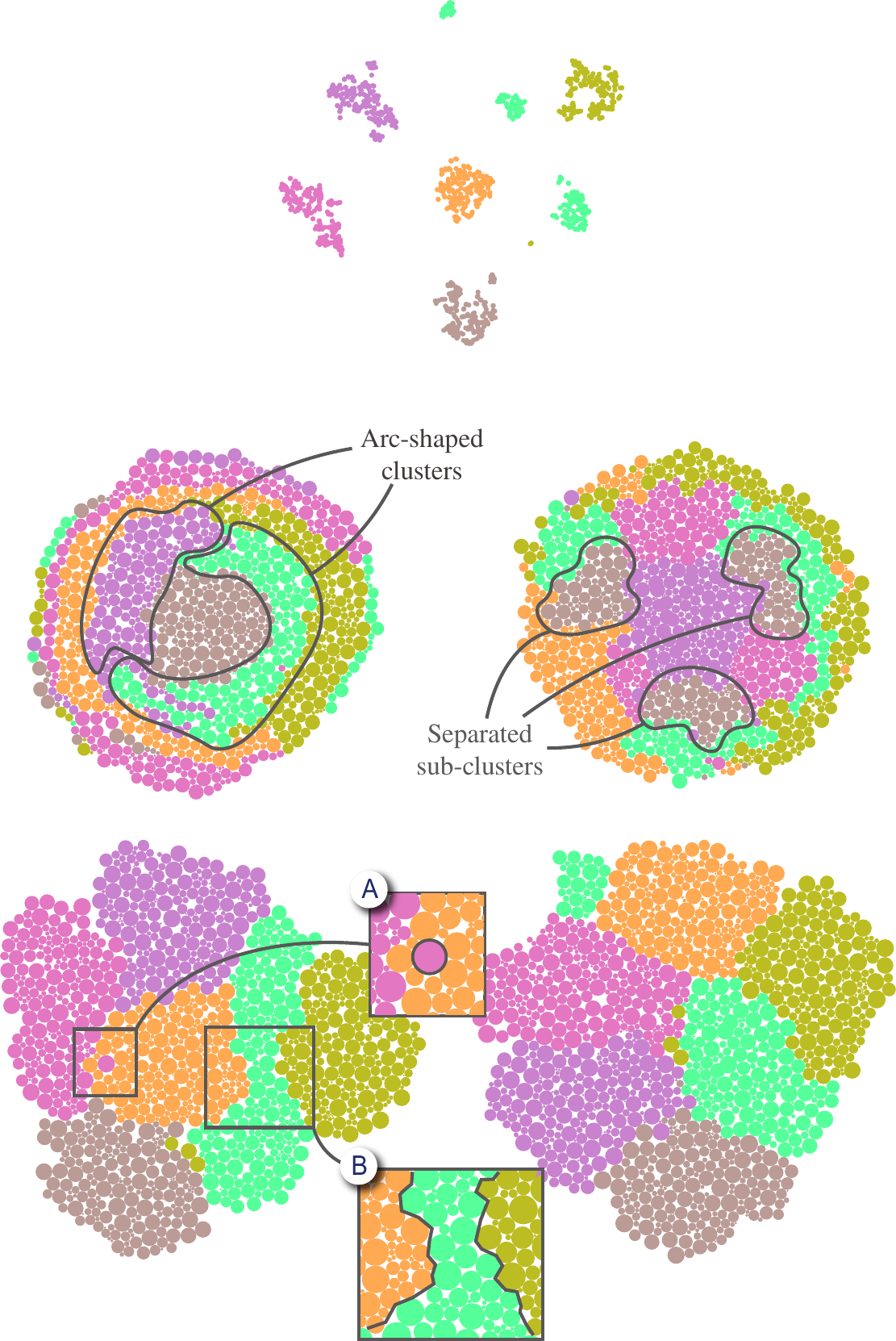}
\put(-160, 255){(a) \emph{t-SNE Projection}}
\put(-210, 136){(b) \emph{SimiFC}}
\put(-70, 136){(c) \emph{AEF}}
\put(-205, 10){(d) \emph{FD}}
\put(-70, 10){(e) \emph{NCP}}
\caption{Comparison of  circle packing results generated by different methods on the MNIST64 dataset.}
\label{fig:compare}
\end{figure}

The results show that \emph{NCP} achieves better convexity and neighborhood preservation without sacrificing compactness: \emph{NCP} obtains an improved solution via the continuation method.
In contrast, \emph{FD} directly optimizes the final objective function, making it prone to becoming trapped in unfavorable local optima, a common pitfall in circle packing optimization.
Specifically, for neighborhood preservation, \emph{NCP} achieved average improvements of 0.042 and 0.070 for $NP_1$, and 0.023 and 0.071 for $NP_2$ under the two cases of radii variances.
Since \emph{NCP} and \emph{FD} use the same planar graph initialization, this gap was caused by the subsequent optimization process.
Unlike \emph{FD}, which optimizes these competing objectives simultaneously, \emph{NCP} utilizes the continuation method to add new objectives progressively, reducing the risk of becoming trapped in local minima in the complex original problem.
\emph{NCP} achieves a similar average compactness to \emph{FD}: both methods adopt the force-directed method as the final step in their optimization processes, which leads to similar compactness.
This conclusion is also supported by the result of \emph{AEF}, which also employs the force-directed method and achieves similar compactness.
\emph{NCP} showed average convexity improvements of 0.004 and 0.015 in the two cases of radii variances.
The main reason  is the parallelized computation strategy in our enhanced power-diagram-based planar graph layout which creates the power diagram for each cluster in a convex sub-region,  providing an initialization in which the clusters have more convex shapes for  subsequent force-directed refinement.
Therefore, it is easier for \emph{NCP} to achieve better convexity.

We also visually compared the circle packing results of these methods to provide a more intuitive explanation of their differences.
Full results for all datasets are provided in the supplemental material.
\cref{fig:compare} shows the results for MNIST64 where the circle radii have a large variance.
\emph{SimiFC} generates arc-shaped clusters (\cref{fig:compare}(b)), and \emph{AEF} may separate a cluster into several sub-clusters (the three sub-clusters of brown circles in \cref{fig:compare}(c)).
Compared to the projection result (\cref{fig:compare}(a)), these two layout results distort the cluster structures and hinder cluster analysis.
In contrast, \emph{FD} (\cref{fig:compare}(d)) and \emph{NCP} (\cref{fig:compare}(e)) well preserve the cluster structures  and their relative positions, making clusters easy to recognize.
However, \emph{FD} displaces circles more frequently.
In \cref{fig:compare}A, one pink circle with thick borders is incorrectly placed inside the cluster of orange circles.
This displacement could lead to incorrect identification of outliers.
Instead, \emph{NCP} correctly places these circles in their respective clusters.
Furthermore, \emph{FD} creates irregular and concave boundaries between clusters, such as the boundaries of the light green cluster in \cref{fig:compare}B.
In contrast, \emph{NCP} forms smooth and convex boundaries between the clusters.

\subsubsection{Running Time}
We evaluated the running time of different circle packing methods on a desktop PC with a 3.00 GHz Intel i9-13900K CPU.
We report  results averaged over five trials and two cases of radii variances per dataset to reduce randomness.
\cref{tab:time} shows that \emph{SimiFC} was the fastest since it employed a simple greedy strategy to place circles.
However, it does not produce satisfactory packing results, as reported in \cref{tab:performance}.
The other three methods take longer time because they incorporate a force-directed method.
Among them, \emph{NCP} is the fastest and generates circle packing results for a dataset with about $1,000$ data items in $2$ seconds.
Its high efficiency is due to the integration of the power-diagram-based method, which provides a good starting point for the force-directed method and thus leads to faster convergence ($1,250$ iterations vs. $10,000$ iterations for \emph{FD}).

\begin{table}[t]
    \caption{Comparison of running time (in seconds).}
    \label{tab:time}
    \centering
\setlength\tabcolsep{4pt}
    \begin{tabular}{l|r|rrrr}
        \toprule
        \textbf{Dataset} & \textbf{Size} & SimiFC & AEF & Ours-FD & Ours-NCP \\
        \midrule
        Boston & 155 & 
        ${0.004}$ & 
        $0.717$ & 
        $4.073$ &
        $0.212$\\
        Dermatology & 259 & 
        ${0.010}$ &
        $1.509$ &
        $6.376$ &
        $0.480$\\
        Ecoli & 336 & 
        ${0.018}$ &
        $2.161$ &
        $9.598$ &
        $0.534$\\
        ExtYaleB & 320 & 
        ${0.018}$ &
        $1.745$ &
        $9.399$ &
        $0.760$\\
        MNIST64 & 1,083 & 
        ${0.219}$ &
        $12.108$ &
        $17.587$ &
        $2.002$\\
        Olive & 572 &
        ${0.059}$ &
        $3.352$ &
        $9.744$ &
        $0.965$\\
        Weather & 366 & 
        ${0.023}$ &
        $2.404$ &
        $8.906$ &
        $0.706$\\
        World12D & 151 & 
        ${0.004}$ &
        $0.782$ &
        $2.761$ &
        $0.349$\\
        \bottomrule
    \end{tabular}
\end{table}

\begin{figure}[t]
\centering
\includegraphics[width=\linewidth]{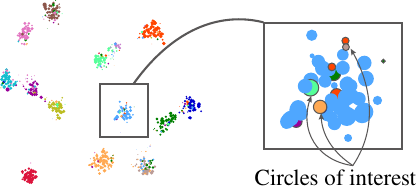}
\caption{Visualizing the Clothing dataset with a scatterplot.}
\label{fig:clothes-scatter}
\end{figure}

\subsection{Use Cases}

We showcase the application of \emph{NCP} to data analysis using two real-world datasets: Clothing~\cite{xiao2015learning} and Boston Housing~\cite{harrison1978hedonic}.

\subsubsection{Clothing}
\label{subsubsec:clothing}
In this use case, we illustrate how \emph{NCP} can be applied to identify and analyze label noise in an image classification dataset.
We used the Clothing dataset, which contains 14 categories of clothing images, with $38.5\%$ of them reported as mislabeled.
Alice is a graduate student.
She aimed to identify and analyze the label noise in this dataset.
To achieve this, Alice trained a ResNet-50 model (accuracy: $76.4\%$) on this dataset for feature extraction and similarity calculation.
Subsequently, she randomly sampled $700$ images for further analysis.
Based on her prior knowledge, Alice focused on two types of label noise. 
The first type is \emph{random noise}, which is often introduced by spammer annotators and is, therefore, irrelevant to the image content.
For example, an image of `sweater' is mislabeled as `shawl', even though there are clear visual differences between the two categories.
Despite such mislabeling, these images can usually be correctly predicted as their ground-truth labels with high confidence scores.
The second type is \emph{content-ambiguity-related noise}.
They mainly come from images with ambiguous content that is hard to categorize, such as the images of `knitwear' and ``sweater.''
These images are usually predicted with low confidence scores, which also means high uncertainty scores.

\vspace{1mm}
\paragraph{Analyzing random noise.}
Initially, Alice visualized the projection results of the sampled images using a scatterplot (\cref{fig:clothes-scatter}).
She used colors to encode their annotated labels and radii to encode the confidence scores,
a larger radius indicating a higher score.
This  facilitated the identification of images with annotated labels that  differ from their neighbors and have high prediction confidence scores.
Alice observed a cluster mainly consisting of light blue circles, while it also included a few circles of different colors.
However, the circles overlapped  each other, causing visual clutter which hindered the analysis.
To address this issue, she then employed the \emph{NCP} method to generate a neighborhood-preserving circle packing result with non-overlapping circles that facilitated sample-level analysis (\cref{fig:clothes}(a)).
The result preserved the cluster structures well, as  circles of the same color are mostly grouped together.
Alice turned her attention to the previously identified cluster and easily identified those large circles of different colors placed inside this cluster.
She suspected that they represented images with random noise because their confidence scores were high.
After examining the associated images, Alice confirmed that several images of `T-shirt' were mislabeled as other labels, such as `knitwear', `shirt', `vest', and `underwear' (\cref{fig:clothes}(a)).
She also checked other regions with circles that showed similar visual patterns and identified more images with random noise.
For example, in region B, a brown circle is placed inside the cluster of other blue circles.
It represents a `windbreaker' image mislabeled as a `suit'.
Similarly, in region C, an image of a `shirt' is mislabeled as a `jacket' (purple circle), placed inside the `shirt' cluster  (orange circles).

\paragraph{Analyzing content-ambiguity-related noise}.
In \cref{fig:clothes}(a),  circles representing images with low confidence scores are hard to recognize due to their small radii.
Therefore, it is hard to identify  images with content-ambiguity-related noise since their confidence scores are usually low.
Alice then changed the encoding and used radii to encode the uncertainty scores, which effectively highlights those ambiguous samples.
Since these images are typically located on the decision boundaries between different predictions, Alice used colors to encode their predicted labels to facilitate the identification of these samples (\cref{fig:clothes}(b)).

First, Alice analyzed region D in \cref{fig:clothes}, where there were large circles on the boundary between the `knitwear' and `sweater' clusters.
This indicates the model's difficulty in predicting the correct labels for these images.
Upon inspection, Alice identified several cases where the model's predictions were inconsistent with the annotated labels.
Amongst them, some images were correctly labeled but mispredicted by the model.
For example, D$_1$ is labeled correctly as a `sweater' , but the model mispredicts it as `knitwear'.

Second, Alice moved to a complex case, region E, characterized by many large circles from three clusters.
The corresponding images belonged to three categories, `windbreaker', `down coat', and `jacket'.
Some were predicted inconsistently with their annotated labels due to  similarity of appearance.
For example, both E$_1$ and E$_2$ are a `down coat' but are mislabeled as a `windbreaker'.
E$_1$ is incorrectly predicted as a `jacket','' while E$_2$ is predicted as ground-truth label `windbreaker'.
Alice noticed that the content-ambiguity-related noise within these categories led to many inconsistent predictions with the annotated labels and thus hindered the model from correctly classifying such images.

\subsubsection{Boston Housing}
\begin{figure}[t]
\centering
\includegraphics[width=\linewidth]{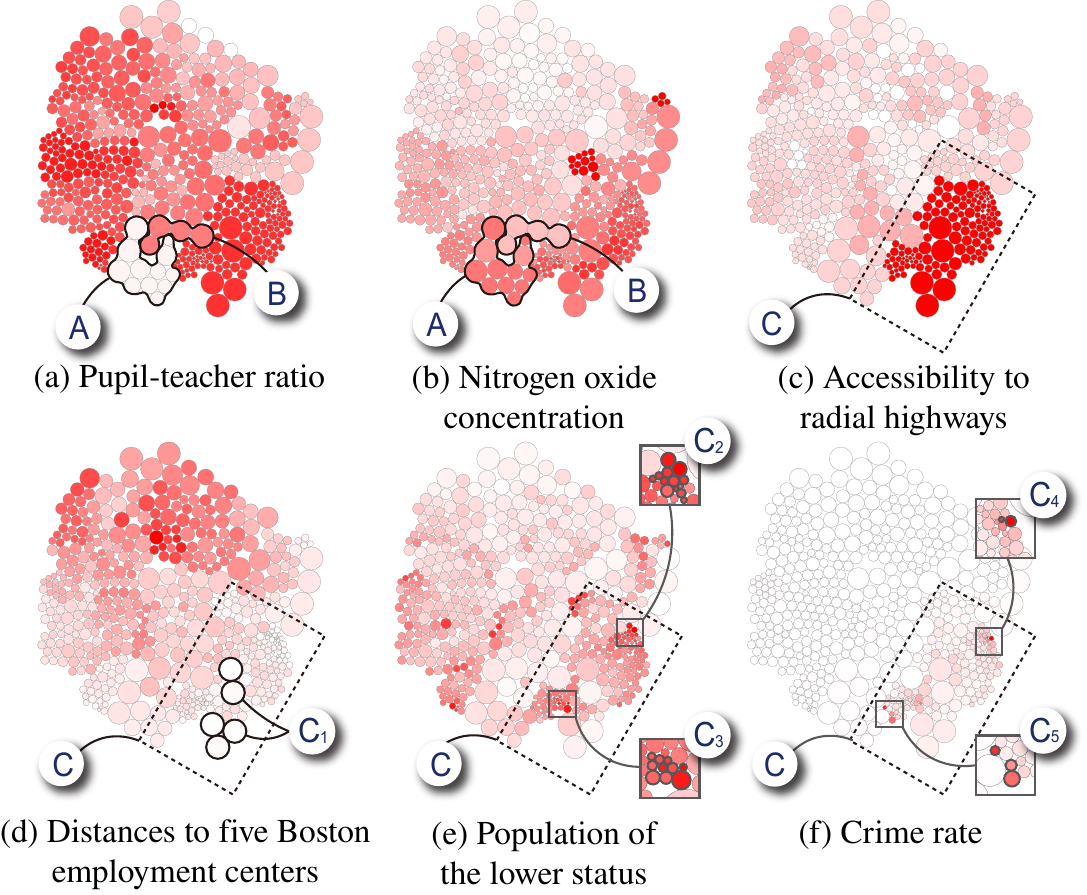}
\caption{Circle packing results for the Boston Housing dataset. Larger circles represent higher housing prices; darker colors indicate higher values for the selected quantitative attribute.}
\label{fig:housing}
\end{figure}

In the second use case, we illustrate how \emph{NCP} can be applied to analyze tabular datasets.
We used the Boston Housing dataset,  containing 506 samples.
Each sample represents a town associated with its housing price along with 13 quantitative attributes derived from multiple factors that affect the housing price, such as educational resources and the environment.
Following previous works~\cite{adetunji2022house, ghatnekar2021explainable}, these quantitative attributes were normalized to derive the features of each sample and calculate their similarity. 
Bob is a real estate agent.
He wanted to analyze how different factors affect housing prices and compare housing prices in similar towns.
To this end, he employed \emph{NCP} to generate a circle packing that placed  similar towns together, simultaneously encoding the housing price and a selected quantitative attribute using  different channels: color and radius (\cref{fig:housing}).
Here, a larger radius corresponds to a higher housing price, and a darker color indicates a higher value for the selected attribute.

Initially, Bob wanted to explore how educational resources affected housing prices.
Therefore, he used  color to encode `pupil-teacher ratio'.
A lighter color represents a lower ratio, indicating more educational resources.
He quickly identified a region where several large, light-colored circles were gathered (region A in \cref{fig:housing}(a)).
This finding suggested that abundant educational resources contributed to  high housing prices.
Adjacent to this region, Bob noted another group of large circles with darker colors (region B in \cref{fig:housing}(a)).
He was curious why these towns had fewer educational resources but still maintained high housing prices compared to those in region A.
To explore this, he encoded other quantitative attributes using colors.
He then analyzed the color distributions of these attributes in regions A and B.
This analysis revealed a clear difference in the distribution of `nitrogen oxide concentrations'.
As shown in \cref{fig:housing}(b), the towns in region B had lighter colors than those in region A, which indicated lower pollution levels.
This suggested that the towns in region B had a better natural environment, which played a key role in their high housing prices.
Such findings are difficult to obtain through  statistical correlations alone, as they involve only a small subset of samples.
In contrast, by visually placing similar houses together and encoding housing prices as circle radii, these findings are more readily identifiable.

Subsequently, Bob wanted to explore the influence of transportation on housing prices.
To this end, he encoded `accessibility to radial highways' using colors.
He observed that the circles in region C had the highest values (\cref{fig:housing}(c)).
However, the large variance in their radii suggested the presence of other factors that affected their housing prices.
To investigate these factors, Bob encoded other quantitative attributes using colors and examined their distributions within this region.
The large circles were associated with shorter `distances to five Boston employment centers' (C$_1$ in \cref{fig:housing}(d)).
This factor also contributes to their high housing prices.
On the contrary, the small circles are associated with larger `populations with lower status' (C$_2$, C$_3$ in \cref{fig:housing}(e)) or higher `crime rates' (C$_4$, C$_5$ in \cref{fig:housing}(f)), which account for their low housing prices.\looseness=-1

\newcommand{\clusterhead}{\emph{Task 1---Cluster Identification}}
\newcommand{\outlierhead}{\emph{Task 2---Outlier Identification}}
\newcommand{\attributehead}{\emph{Task 3---Quantitative Attribute Comparison}}

\newcommand{\myquote}[1]{\emph{``#1''}}

\newcommand{\clusterbody}{identify the cluster}
\newcommand{\outlierbody}{identify the outlier}
\newcommand{\attributebody}{compare quantitative values}

{
\subsection{User Study}
In addition to the quantitative evaluation and the use cases, we also conducted a user study to demonstrate the effectiveness and usefulness of NCP in data analysis.
The study used 9 real-world datasets: 8 used in Sec.~\ref{subsec:quantitative} and 1 used in Sec.~\ref{subsubsec:clothing}.

\subsubsection{Study Setup}
\paragraph{Participants}
We recruited 16 participants (P1–P16), aged 24 to 32 years, comprising graduate students and professors with extensive experience ($\ge$2 years) in visual analytics and information visualization. 
Upon completion, each participant received  \$15 compensation, regardless of performance.

\begin{figure*}[t]
\centering
\includegraphics[width=\linewidth]{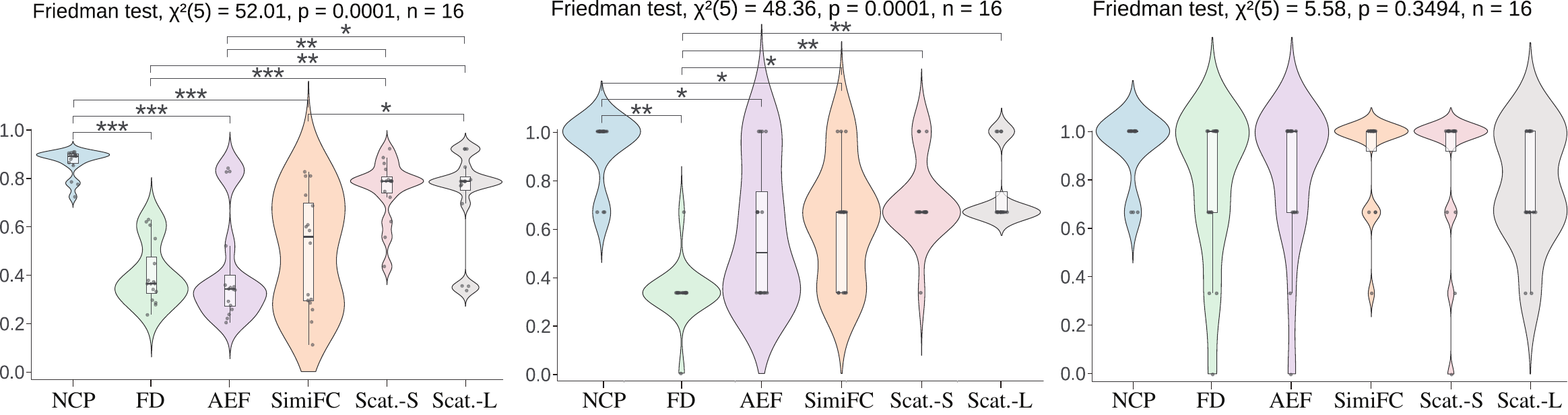}
\put(-415, -10){(a)}
\put(-248, -10){(b)}
\put(-75, -10){(c)}
\caption{User study results for three tasks: (a) \clusterhead{}; (b) \outlierhead{}; (c) \attributehead{}. Here, * indicates $p< 0.05$, ** indicates $p< 0.01$, *** indicates $p< 0.001$. Scat.-S stands for Scatter-S, and Scat.-L stands for Scatter-L.}
\label{fig:userstudy-acc}
\end{figure*}


\paragraph{Baseline methods}
We used 5 baseline methods: 3 circle packing methods (\emph{SimiFC}, \emph{AEF}, and \emph{FD}) used in Sec.~\ref{subsec:quantitative}, and 2 scatterplot methods (\emph{Scatter-S} and \emph{Scatter-L}) that allow overlap between circles.
In the scatterplot methods, circle positions were determined by t-SNE, consistent with the initialization results of our method.
Scatter-S uses the \emph{size} of circles to encode quantitative attributes, while Scatter-L uses color \emph{lightness}.
These two methods are widely used in data analysis~\cite{munzner2014visualization}.

\paragraph{Task Design}
Our study consisted of three tasks commonly used in visual analytics~\cite{brehmer2013tasks,yang2024foundation,yang2023survey}.

\clusterhead{}: Participants were required to identify the cluster to which a highlighted circle belonged.
They could use a lasso or click to select the circles in the identified cluster and submit their answer.

\outlierhead{}: Participants were required to judge whether a highlighted circle was an outlier.
They could analyze the neighborhood relationships around the highlighted circle before submitting their answer.

\attributehead{}: Participants were required to compare three highlighted circles and identify the one corresponding to the largest quantitative value.
For \emph{Scatter-L}, this was achieved by comparing the lightness of their colors, while for the other methods, it was done by comparing their sizes.

\paragraph{Study Protocol}
Participants started by signing a consent form and watching a tutorial video about the study procedure, system interactions, and tasks.
Following a within-subjects design, each participant was required to evaluate six different methods and finish all three tasks sequentially.
Each task consisted of both a practice and a test session.
In the practice session, participants were required to finish six trials, one for each method.
After completing the practice session and confirming that they fully understood the tasks and methods, participants proceeded to the test session.
Participants were allowed to take short breaks whenever they requested one.
Upon completing each task, we assessed the participants’ workloads and fatigue levels using NASA’s Task Load Index~\cite{sandra2006nasa} with brief descriptions of the six methods, and collected their feedback on the methods.
For each trial, we recorded  participants' answers and completion times.

To control the experiment duration and reduce the learning effect, the nine datasets were evenly distributed across three tasks, with each task assigned three datasets.
Thus, each participant completed 54 trials (3 tasks $\times$ 3 datasets $\times$ 6 methods). 
The entire study lasted 45-60 minutes.
The method order was also counterbalanced to reduce the learning effect.

\subsubsection{Result Analysis}
We analyzed both task accuracy  and participants' subjective ratings for workload and fatigue.

\paragraph{Accuracy} For each task, we computed participants' average accuracy across different methods.
We conducted Friedman tests and pair-wise Wilcoxon signed-rank tests with Bonferroni correction for multiple comparisons.
Statistical test results are shown in Fig.~\ref{fig:userstudy-acc}.
The Friedman test results indicate significant differences between methods in the first two tasks: \clusterbody{} ($\chi^2(5)=52.01$, $p < 0.0001$), \outlierbody{} ($\chi^2(5)=48.36$, $p < 0.0001$), and no significant difference in \attributebody{} ($\chi^2(5)=5.58$, $p = 0.3494$).
In the subsequent analysis, we focus on pairwise comparisons.

\begin{figure}[t]
\centering
\includegraphics[width=\linewidth]{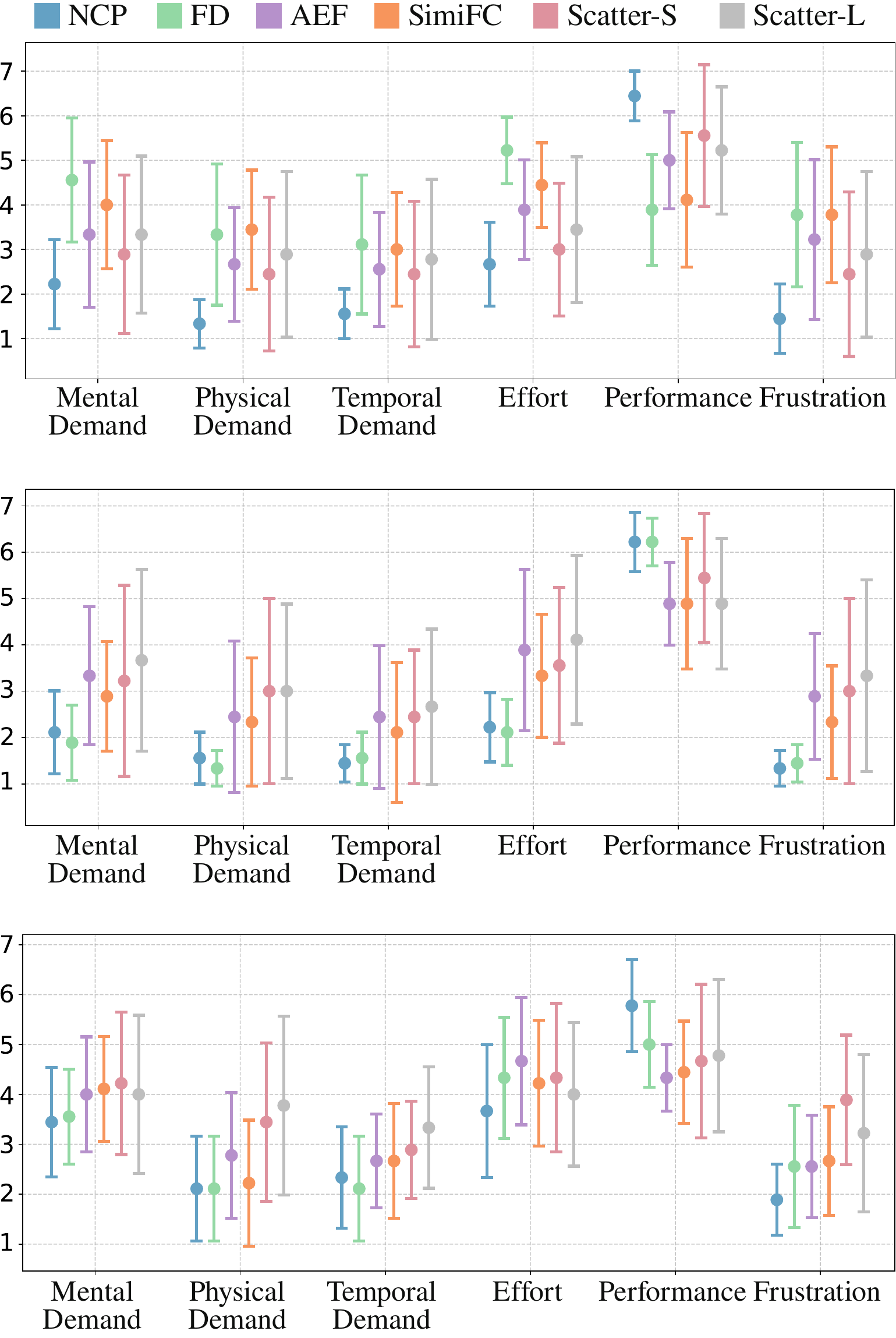}
\put(-120, 230){(a)}
\put(-120, 110){(b)}
\put(-120, -10){(c)}
\caption{Participants' workload and fatigue levels according to NASA's Task Load Index: (a) \clusterhead{}; (b) \outlierhead{}; (c) \attributehead{}.
  Here, error bars show the 95\% confidence intervals.}
\label{fig:userstudy-effort}
\end{figure}

\clusterhead{}: \emph{NCP} delivers superior performance  to \emph{SimiFC}, \emph{AEF}, and \emph{FD} while performing comparably to \emph{Scatter-S} and \emph{Scatter-L}.
Participants consistently praised NCP for its ability to produce \myquote{clear cluster boundaries.}
This is because \emph{NCP} explicitly optimizes cluster convexity during force-directed refinement.
Comparing performance between \emph{NCP} and scatterplot methods (\emph{Scatter-S} and \emph{Scatter-L}) indicates that reducing the space between clusters does not hamper the identification of clusters but enhances layout compactness.
This optimized use of space improves visual clarity and thus facilitates deeper and more effective data analysis.

\outlierhead{}: \emph{NCP} significantly outperforms \emph{SimiFC}, \emph{AEF}, and \emph{FD}, while performing comparably to \emph{Scatter-S} and \emph{Scatter-L}.
Participants described outliers in \emph{NCP} as \myquote{clearly noticeable.}
Compared to other circle packing methods, \emph{NCP} achieves better cluster convexity and clearer cluster boundaries, making outliers that deviate from clusters with the same class label more noticeable.
Compared to \emph{Scatter-S} and \emph{Scatter-L}, the lack of overlap and compactness of \emph{NCP} is a double-edged sword.
On the one hand, it ensures that every circle is clearly visible, preventing outliers from being too small to notice or occluded by others.
On the other hand, the compactness reduces the space between circles, potentially drawing outliers that originally deviated from clusters back toward the cluster boundaries, making some of them harder to identify as outliers.

\attributehead{}: Analysis results show no significant difference between  methods.
However, \emph{NCP} achieved the highest mean accuracy (0.933) with the lowest standard deviation (0.138), highlighting its effectiveness in helping users to compare attribute values.

\paragraph{Workload} 
Fig.~\ref{fig:userstudy-effort} shows participant workload and fatigue levels measured using NASA's Task Load Index, including mental demand, physical demand, temporal demand, effort, performance, and frustration.
For all three tasks, \emph{NCP} performed better than other circle packing methods across all six measures, with lower mental and physical demands, reduced effort, less temporal pressure, higher performance, and lower frustration.
P3 commented that NCP could \myquote{balance the cluster overview and local details well}, enhancing analysis efficiency and reducing workload.
When compared to scatterplot methods, \emph{Scatter-S} and \emph{Scatter-L} resulted in greater physical demands and higher frustration, particularly in \attributehead{}.
This is because scatterplot methods with lower compactness have reduced space efficiency, leading to poorer readability.
P5 commented: \myquote{At the beginning, I could hardly compare the highlighted circles since they were too small.}
Even with the provided zooming in and out interactions, P7 noted: \myquote{I have to zoom in several times}, which caused a loss of broader context.}
In contrast, the lack of overlap and compactness of \emph{NCP} significantly enhance readability, enabling participants to locate their targets and perform comparisons with greater ease and efficiency.


\section{Expert Feedback and Discussion}

We interviewed three experts (E$_1$--E$_3$) specializing in visual analytics and machine learning.
Each interview took approximately an hour, including a 10-minute introduction to \emph{NCP}, a 30-minute session where we presented our use cases and collected expert feedback, and a 20-minute discussion.
Overall, the experts highlighted the utility of \emph{NCP} in data analysis and its convenient integration with existing algorithms.
We also identified several potential directions for future research based on the interviews.

\begin{figure}[t]
\centering
\includegraphics[width=\linewidth]{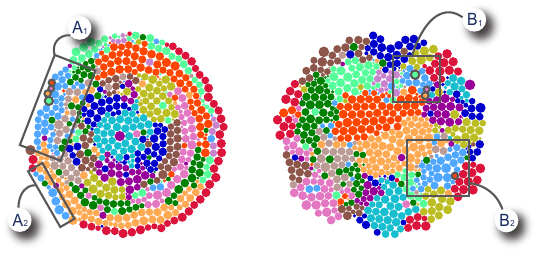}
\put(-215, -5){(a) \emph{SimiFC}}
\put(-85, -5){(b) \emph{AEF}}
\caption{Two alternative layouts for the Clothing dataset.}
\label{fig:clothes-alt}
\end{figure}

\subsection{Usability}

\vspace{1mm}
\paragraph{Enhancing data analysis}
Overall, the experts acknowledged the capability of \emph{NCP} to facilitate the analysis and comparison of similar data items of interest.
They commented that they could easily understand the cluster structures and identify outliers with \emph{NCP}.
For example, in \cref{fig:clothes}(a), the cluster structures are clearly shown since circles of the same colors are mostly grouped together.
Circles representing mislabeled images stand out due to the  color differences from their neighbors.
This prompts users to examine whether they are mislabeled.
The experts also noted that the convexity of the cluster shapes was beneficial for understanding clusters, and consequently for identifying outliers efficiently.

The experts were also asked to compare the circle packing results generated by \emph{NCP} and the baseline methods, \emph{SimiFC} and \emph{AEF}.
They agreed that all these methods could generate compact circle packing results, but the visual patterns differed a lot in each case.
Generally, they favored \emph{NCP} and noted that its results preserved cluster structures well.
In contrast, the experts identified distortions of clusters in the results of the baseline methods.
Using the clothing dataset from the first use case as an example,
\emph{SimiFC} generated arc-shaped clusters, as shown in \cref{fig:clothes-alt}(a).
E$_3$ pointed out that these visual patterns might lead to misunderstandings of the data relationships.
For example, users may perceive a certain sequential order in the placement of circles from the inside to the outside.
Both \emph{SimiFC} and \emph{AEF} might separate the clusters into different sub-clusters.
For example, the cluster of light blue circles is disjoint in \emph{SimiFC} (Figs.~\ref{fig:clothes-alt}A$_1$ and A$_2$) and \emph{AEF} (Figs.~\ref{fig:clothes-alt}B$_1$ and B$_2$).
In addition, the green, orange, pink, and red circles representing the samples with random noise are placed near the center of  \cref{fig:clothes}A.
However,  in Figs.~\ref{fig:clothes-alt}A$_1$, B$_1$, and B$_2$, they are on the boundary of the cluster consisting of light blue circles.
This hinders the identification of these outliers.

\paragraph{Pipeline flexibility}
While our NCP method is primarily designed for visualization, its flexible optimization pipeline enables potential applications to other domains by supporting diverse algorithms and objective functions.
First, our pipeline supports the integration of different projection methods, power-diagram-based methods, and utilized forces. 
This flexibility allows users to tailor their choices in each phase according to their specific needs.
For example, in the case of textual data analysis, users can employ least-squares projection in the planar graph initialization, which has been shown to be effective in preserving neighborhood relationships in textual data~\cite{paulovich2008hipp}.
Second, the continuation method can be extended to accommodate more objectives, such as ensuring efficient connectivity while minimizing interference in wireless sensor network design, and enhancing the aesthetic arrangement of circles in design and art applications.
By carefully determining the sequence of objectives and constraints to be incorporated into the optimization process, it achieves a well-balanced integration of multiple criteria.
This adaptability improves the effectiveness of our method in a variety of optimization contexts, making it applicable to other domains.
\subsection{Limitations and Future Work}

\paragraph{Dynamic parameter tuning}
Our \emph{NCP} method achieves a balance between three optimization objectives using the parameters $\alpha$ and $\beta$.
In our implementation, we determined their values using a grid search, which is time-consuming.
We would like to explore automatic parameter tuning methods to ease this process, such as multi-task learning~\cite{liu2019loss}.
In addition, E$_2$ pointed out that user preferences for optimization objectives could vary in different real-world applications.
For example, E$_2$ said, ``When the model prediction is reliable, I would prioritize optimizing convexity to enhance the perceptual clarity of clusters. 
Otherwise, I would focus more on neighborhood preservation, which helps me identify prediction errors more easily.''
Therefore, integrating user feedback to dynamically adjust these parameters is essential, to allow users to tailor the circle packing to their specific requirements.

\paragraph{Integration with other visualizations techniques}
Our experts have identified several opportunities to enhance \emph{NCP} through integration with other visualization techniques.
First, they proposed that the layout could be enhanced to provide more guidance for data exploration.
A possible method would be to design informative glyphs displayed within circles, to offer more details of data items.
For example, in the second use case, pie charts can be used to show how different factors contribute to housing prices instead of encoding each quantitative attribute individually.
Another method would be to select representative data items and show them in the empty space, to help users better understand the data.
Second, E$_2$ and E$_3$ suggested employing \emph{NCP} for hierarchical exploration of large data.
By building a hierarchy for the data items, a subset of data items can be sampled and visualized as in the first use case.
Alternatively, circles can represent groups of similar data items instead of individual ones.
Users can navigate the whole dataset with the zooming function.

\paragraph{Interactive circle packing}
User interactions could also be introduced to improve the packing result of \emph{NCP}.
Specifically, E$_3$ expressed the need to author functions to manipulate the relative positions of circles.
``If I notice that some circles should form a cluster but are not placed adjacently, I want to adjust their positions to make this cluster more evident.''
We should consider allowing users to move a few representative circles using drag-and-drop.
During this process, the proximity between circles changes.
As a result, further investigation is required to ensure the stability of circle packing and maintain users' mental maps.
We also should consider supporting users  directly specifying must-link constraints between data items and incorporating these constraints into the optimization process, to ensure that the generated circle packing  places circles of interest adjacent to one another.


\section{Conclusions}

\label{sec:conclusion}

In this paper, we have developed a new layout method, \emph{NCP}, for generating a neighborhood-preserving non-uniform circle packing.
We formulate circle packing as a planar graph embedding problem and solve it using the continuation method. 
By progressively incorporating multiple optimization objectives and constraints, this method steers the optimization towards a more favorable solution.
Our quantitative comparison to baselines shows that \emph{NCP} performs better in terms of neighborhood preservation and convexity while achieving comparable compactness.
Two use cases further demonstrate its application in data analysis.

\appendix
\subsection{Appendix A: Projection Method and  Associated Parameters}
We conducted quantitative experiments to select the best projection method and  associated parameters for the neighborhood-preserving planar graph initialization.

\subsubsection*{Experimental Setting}
\paragraph{Datasets}
We employed eight real-world datasets used in Xia~\etal~\cite{xia2021revisiting}, including Boston~\cite{Sedlmair2012Taxonomy}, Dermatology~\cite{Dua2017UCI}, Ecoli~\cite{Dua2017UCI}, ExtYaleB~\cite{Georghiades2001from}, MNIST64~\cite{Dua2017UCI}, Olive~\cite{forina1983classification}, Weather~\cite{ventocilla2020comparative}, and World12D~\cite{Sedlmair2012Taxonomy}.

\paragraph{Measure}
We used the neighborhood preservation degree~\cite{kruiger2017graph, zhong2023force} to measure how well the neighborhood relationships in the high-dimensional data space are preserved in the projection results.

\paragraph{Methods}
We identified five projection methods: t-SNE~\cite{van2008visualizing}, UMAP~\cite{mcinnes2018umap}, PCA~\cite{wold1987principal}, MDS~\cite{kruskal1978multidimensional}, and NMF~\cite{lee1999learning}, based on previous studies~\cite{vernier2020quantitative, xia2021revisiting}.

\paragraph{Parameters}
We explored \emph{perplexity} of t-SNE and \emph{neighbors} of UMAP, which are key factors that affect the neighborhood preservation of these methods.
The candidate parameters were set the same as  in Xia~\etal~\cite{xia2021revisiting}.
Specifically, the \emph{perplexity} of t-SNE was chosen in $\{5, 15, 30, 40, 50\}$, and the \emph{neighbors} of UMAP was chosen in $\{4, 7, 10, 13, 16\}$.
PCA, MDS, and NMF applied  default parameter settings.

\subsubsection*{Results}
\begin{figure*}[tp]
\centering
\includegraphics[width=0.9\linewidth]{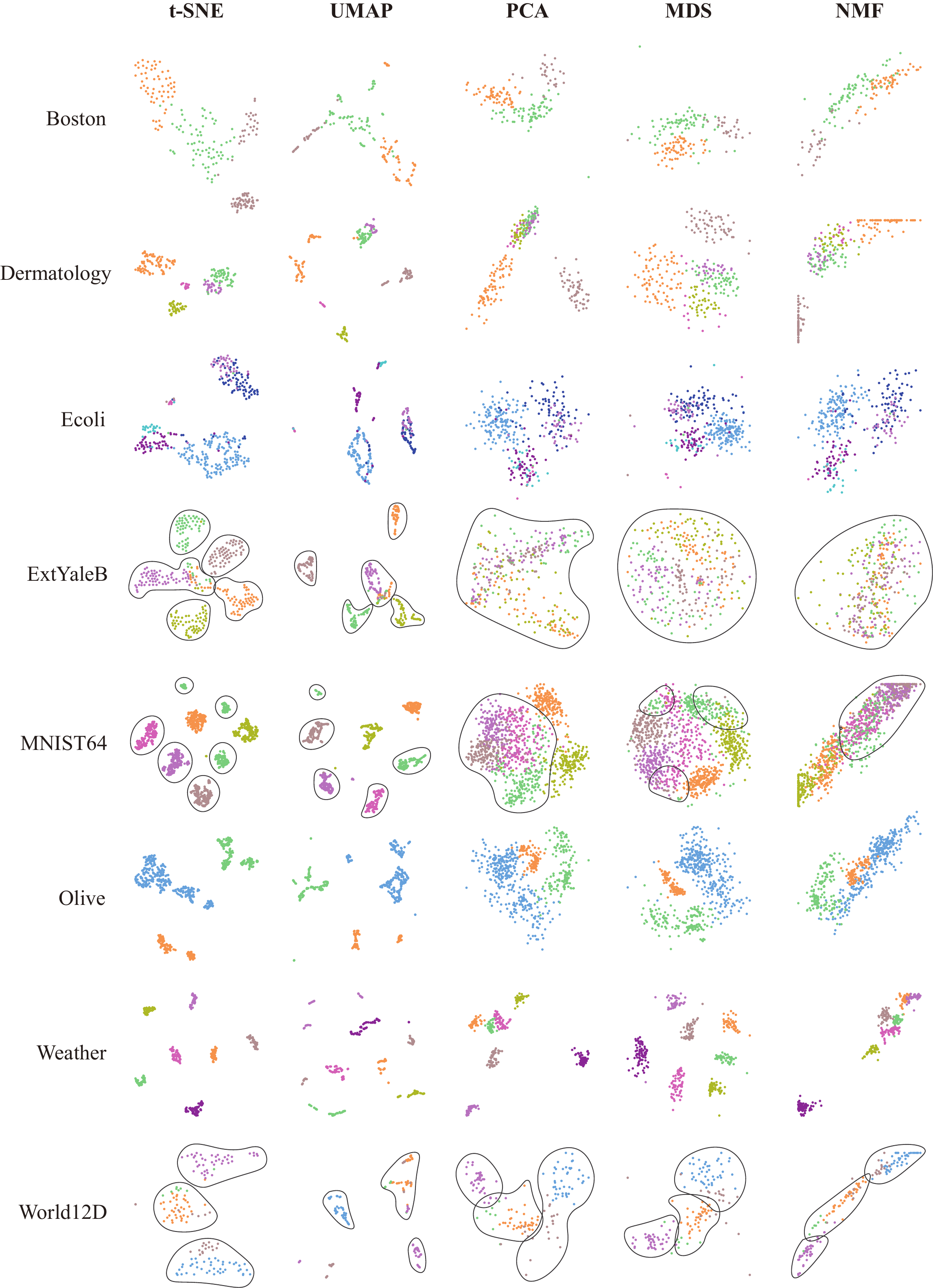}
\caption{Projection results for the eight datasets.}
\label{fig:sidebyside-proj}
\end{figure*}

For t-SNE and UMAP, we first identified the best parameter settings based on the average neighborhood preservation across the eight datasets.
The perplexity of t-SNE was set to $15$, and the Neighbors of UMAP was set as $4$.
We then compared the neighborhood preservation for different datasets.
As shown in \cref{tab:method-comparison}, t-SNE and UMAP significantly outperformed the other methods, with t-SNE outperforming UMAP on seven datasets and on average (0.392 v.s. 0.351).
Fig.~\ref{fig:sidebyside-proj} provides a visual comparison between these methods.
In datasets like ExtyaleB and MNIST64, t-SNE and UMAP clearly separated different samples with different labels, but the other three methods tended to confuse them.
This explains the clear gap between t-SNE/UMAP and the other methods.
When comparing the results of t-SNE and UMAP, the points in t-SNE are more evenly distributed compared to UMAP, which better reflects the neighborhood relationships.
Therefore, we selected t-SNE with a \emph{perplexity} of 15 in our layout method.

\begin{table}[t]
    \centering
    \caption{Neighborhood preservation comparison for different projection methods. }
    \label{tab:method-comparison}

    \begin{tabular}{l|cccccc}
       \toprule
        Dataset & t-SNE & UMAP & PCA & MDS & NMF \\
        \midrule
        Boston &
        \textbf{0.374} & 0.346 & 0.184 & 0.230 & 0.167   \\
        
        Dermatology & 
       \textbf{0.366} & 0.328 & 0.115 & 0.151 & 0.087 \\ 
        
        Ecoli & 
         \textbf{0.380} & 0.338 & 0.159 & 0.186 & 0.146   \\ 

        ExtYaleB &
        \textbf{0.474} & 0.373 & 0.109 & 0.200 & 0.101   \\
        
        MNIST64 &
         \textbf{0.346} & 0.288 & 0.074 & 0.082 & 0.039  \\
        
        Olive &
         \textbf{0.375} & 0.335 & 0.165 & 0.177 & 0.143  \\
        
        Weather &
        \textbf{0.473} & 0.400 & 0.208 & 0.263 & 0.199   \\
        
        World12D &
        0.350 & \textbf{0.400} & 0.280 & 0.313 & 0.240   \\ 
        \midrule
        
        Average &
         \textbf{0.392} & 0.351 & 0.162 & 0.200 & 0.140  \\
        \bottomrule
        
    \end{tabular}
    
\end{table}



\subsection*{Appendix B: Comparison of Different Initialization Methods}
{We also conducted a study to compare  circle packing results generated using different initialization methods.
We used the same datasets and metrics as those in Sec.~5.1.}

\paragraph{Results}
The experimental results are presented in Table~\ref{tab:performance-initial}.
Overall, t-SNE achieved the best average performance in terms of neighborhood preservation.
This finding aligns with results in Appendix~A, supporting the importance of initialization methods in neighborhood preservation.
All  methods performed comparably in terms of compactness and convexity.
This is because these objectives are incorporated into optimization in subsequent steps, which are identical for all  methods reported in this section.

{We also visually compared the circle packing results of the reported methods in Figs.~\ref{fig:sidebyside-init1} and~\ref{fig:sidebyside-init2} to provide a more intuitive explanation of their differences.
Overall, all methods were able to form reasonable clusters, as indicated by their comparable convexity.
However, there were differences in  neighborhood preservation within clusters.
Taking the MNIST64 dataset in Fig.~\ref{fig:sidebyside-init1} as an example, two highly similar digits `1' were correctly placed together in the results of UMAP and t-SNE, whereas in the results of other methods, they were separated.
This separation complicates the accurate interpretation of proximity within clusters.

\begin{table*}[t!]
    \centering
    \caption{Performance comparison in terms of neighborhood preservation ($NP_1$, $NP_2$), compactness, and convexity.
    }\label{tab:performance-initial}
    
   \begin{subtable}[t]{\linewidth}
   \captionsetup{font={normal}}
\setlength\tabcolsep{2pt}
   \resizebox{\textwidth}{!}{
    \begin{tabular}{l|ccccccccccccccccccccc}
       \toprule
        \multirow{3}{*}{\textbf{Dataset}}
        & \multicolumn{5}{c}{${NP_1}$}
        & \multicolumn{5}{c}{${NP_2}$}
        & \multicolumn{5}{c}{\emph{Compactness}}
        & \multicolumn{5}{c}{\emph{Convexity}} \\
        \cmidrule(lr){2-6} \cmidrule(lr){7-11} \cmidrule(lr){12-16} \cmidrule(lr){17-21}
        & \multicolumn{4}{c}{Baseline} & {Ours}
        & \multicolumn{4}{c}{Baseline} & {Ours}
        & \multicolumn{4}{c}{Baseline} & {Ours}
        & \multicolumn{4}{c}{Baseline} & {Ours} \\
        \cmidrule(lr){2-5} \cmidrule(lr){6-6} \cmidrule(lr){7-10} \cmidrule(lr){11-11} \cmidrule(lr){12-15} \cmidrule(lr){16-16} \cmidrule(lr){17-20} \cmidrule(lr){21-21}
        & {NMF} & {PCA} & {MDS} & {UMAP} & {NCP}
        & {NMF} & {PCA} & {MDS} & {UMAP} & {NCP}
        & {NMF} & {PCA} & {MDS} & {UMAP} & {NCP}
        & {NMF} & {PCA} & {MDS} & {UMAP} & {NCP} \\
        \midrule
        Boston &
        0.150 & 0.170 & 0.219 & 0.301 & \textbf{0.341} &
        0.305 & 0.336 & 0.389 & 0.422 & \textbf{0.423} &
        \textbf{0.890} & 0.881 & 0.879 & 0.876 & 0.889 &
        0.810 & \textbf{0.851} & 0.797 & 0.807 & 0.809 \\
        
        Dermatology & 
        0.119 & 0.125 & 0.160 & 0.280 & \textbf{0.311} &
        0.256 & 0.259 & 0.281 & \textbf{0.399} & 0.385 &
        \textbf{0.886} & 0.881 & 0.868 & 0.872 & 0.874 &
        0.795 & 0.771 & 0.802 & \textbf{0.803} & 0.790 \\
        
        Ecoli & 
        0.164 & 0.163 & 0.191 & 0.316 & \textbf{0.344} &
        0.276 & 0.292 & 0.308 & 0.393 & \textbf{0.411} &
        0.870 & \textbf{0.875} & 0.862 & 0.871 & 0.874 &
        0.784 & 0.801 & 0.794 & 0.812 & \textbf{0.831} \\ 

        ExtYaleB &
        0.201 & 0.225 & 0.241 & 0.387 & \textbf{0.422} &
        0.296 & 0.322 & 0.357 & 0.469 & \textbf{0.487} &
        0.881 & 0.868 & 0.865 & 0.870 & 0.868 &
        0.781 & 0.778 & 0.763 & 0.777 & \textbf{0.806} \\ 
        
        MNIST64 &
        0.073 & 0.090 & 0.092 & 0.213 & \textbf{0.306} &
        0.143 & 0.172 & 0.186 & 0.325 & \textbf{0.372} &
        0.871 & 0.860 & 0.861 & 0.869 & 0.863 &
        0.804 & 0.800 & \textbf{0.834} & 0.787 & 0.827 \\
        
        Olive &
        0.147 & 0.157 & 0.159 & 0.269 & \textbf{0.328} &
        0.277 & 0.280 & 0.285 & 0.371 & \textbf{0.414} &
        0.868 & 0.861 & \textbf{0.866} & 0.865 & 0.862 &
        0.825 & 0.824 & \textbf{0.829} & 0.789 & 0.817 \\
        
        Weather &
        0.179 & 0.188 & 0.223 & 0.351 & \textbf{0.448} &
        0.332 & 0.350 & 0.359 & 0.527 & \textbf{0.567} &
        0.868 & 0.864 & \textbf{0.868} & 0.867 & \textbf{0.868} &
        0.811 & 0.797 & \textbf{0.811} & \textbf{0.811} & 0.795 \\
        
        World12D &
        0.235 & 0.312 & 0.289 & 0.319 & \textbf{0.379} &
        0.375 & 0.442 & 0.447 & \textbf{0.483} & 0.463 &
        \textbf{0.901} & 0.883 & 0.881 & 0.876 & 0.896 &
        0.734 & \textbf{0.819} & 0.807 & \textbf{0.828} & 0.809 \\
        
        \midrule
        
        \textbf{Average} &
        0.158 & 0.179 & 0.197 & 0.304 & \textbf{0.360} &
        0.282 & 0.307 & 0.326 & 0.423 & \textbf{0.440} &
        \textbf{0.879} & 0.872 & 0.869 & 0.871 & 0.874 &
        0.793 & 0.805 & 0.805 & 0.802 & \textbf{0.811} \\
        
        \bottomrule
    \end{tabular}
    }
    \caption{Performance in the case where circle radii have a large variance.}
    \end{subtable} 

    \begin{subtable}[t]{\linewidth}
    \captionsetup{font={normal}}
\setlength\tabcolsep{2pt}
    \resizebox{\textwidth}{!}{
    \begin{tabular}{l|ccccccccccccccccccccc}
       \toprule
        \multirow{3}{*}{\textbf{Dataset}}
        & \multicolumn{5}{c}{${NP_1}$}
        & \multicolumn{5}{c}{${NP_2}$}
        & \multicolumn{5}{c}{\emph{Compactness}}
        & \multicolumn{5}{c}{\emph{Convexity}} \\
        \cmidrule(lr){2-6} \cmidrule(lr){7-11} \cmidrule(lr){12-16} \cmidrule(lr){17-21}
        & \multicolumn{4}{c}{Baseline} & {Ours}
        & \multicolumn{4}{c}{Baseline} & {Ours}
        & \multicolumn{4}{c}{Baseline} & {Ours}
        & \multicolumn{4}{c}{Baseline} & {Ours} \\
        \cmidrule(lr){2-5} \cmidrule(lr){6-6} \cmidrule(lr){7-10} \cmidrule(lr){11-11} \cmidrule(lr){12-15} \cmidrule(lr){16-16} \cmidrule(lr){17-20} \cmidrule(lr){21-21}
        & {NMF} & {PCA} & {MDS} & {UMAP} & {NCP}
        & {NMF} & {PCA} & {MDS} & {UMAP} & {NCP}
        & {NMF} & {PCA} & {MDS} & {UMAP} & {NCP}
        & {NMF} & {PCA} & {MDS} & {UMAP} & {NCP} \\
        \midrule
        Boston &
        0.175 & 0.181 & 0.206 & 0.332 & \textbf{0.365} &
        0.309 & 0.345 & 0.366 & \textbf{0.442} & 0.437 & 
        \textbf{0.895} & 0.876 & 0.870 & 0.879 & 0.888 &
        0.819 & 0.824 & \textbf{0.844} & 0.826 & 0.833 \\
        
        Dermatology & 
        0.115 & 0.126 & 0.153 & 0.320 & \textbf{0.364} &
        0.240 & 0.246 & 0.272 & 0.398 & \textbf{0.399} &
        0.875 & \textbf{0.877} & 0.862 & 0.875 & 0.875 &
        0.834 & 0.797 & \textbf{0.860} & 0.828 & 0.845 \\
        
        Ecoli & 
        0.172 & 0.170 & 0.171 & 0.311 & \textbf{0.350} &
        0.273 & 0.278 & 0.297 & 0.404 & \textbf{0.416} & 
        0.869 & \textbf{0.866} & 0.862 & 0.862 & \textbf{0.866} &
        0.806 & \textbf{0.841} & 0.830 & 0.816 & 0.821 \\
        
        ExtYaleB &
        0.194 & 0.220 & 0.216 & 0.385 & \textbf{0.476} &
        0.302 & 0.308 & 0.354 & 0.475 & \textbf{0.505} &
        0.870 & 0.862 & 0.855 & \textbf{0.869} & 0.862 & 
        0.775 & 0.760 & 0.791 & 0.765 & \textbf{0.808} \\   
        
        MNIST64 &
        0.070 & 0.091 & 0.100 & 0.225 & \textbf{0.327} &
        0.138 & 0.165 & 0.185 & 0.326 & \textbf{0.375} & 
        0.853 & 0.852 & 0.850 & \textbf{0.854} & \textbf{0.854} & 
        0.829 & 0.834 & \textbf{0.835} & 0.800 & 0.808 \\ 
        
        Olive &
        0.161 & 0.162 & 0.158 & 0.273 & \textbf{0.342} &
        0.274 & 0.281 & 0.273 & 0.371 & \textbf{0.422} &
        0.861 & 0.857 & 0.855 & \textbf{0.859} & 0.858 & 
        0.819 & 0.848 & \textbf{0.851} & 0.789 & 0.839 \\
        
        Weather &
        0.192 & 0.212 & 0.241 & 0.379 & \textbf{0.483} &
        0.325 & 0.353 & 0.352 & 0.539 & \textbf{0.578} &
        0.877 & \textbf{0.867} & 0.862 & 0.859 & 0.860 &
        0.850 & 0.841 & \textbf{0.847} & 0.805 & 0.816 \\   
        
        World12D &
        0.264 & 0.268 & 0.323 & 0.363 & \textbf{0.380} &
        0.373 & 0.431 & 0.450 & 0.473 & \textbf{0.477} &
        0.881 & \textbf{0.885} & 0.879 & 0.876 & \textbf{0.885} &
        0.768 & 0.800 & 0.808 & \textbf{0.845} & 0.796 \\
        
        \midrule
        
        \textbf{Average} &
        0.168 & 0.179 & 0.196 & 0.323 & \textbf{0.386} &
        0.280 & 0.301 & 0.319 & 0.429 & \textbf{0.451} &
        0.875 & \textbf{0.867} & 0.862 & 0.866 & \textbf{0.868} & 
        0.813 & 0.818 & \textbf{0.833} & 0.809 & 0.825 \\  
        
        \bottomrule
    \end{tabular}
    }
    \caption{Performance in the case where circle radii have a small variance.}
    \end{subtable} 

\end{table*}

\begin{figure*}[tp!]
\centering
\includegraphics[width=0.9\linewidth]{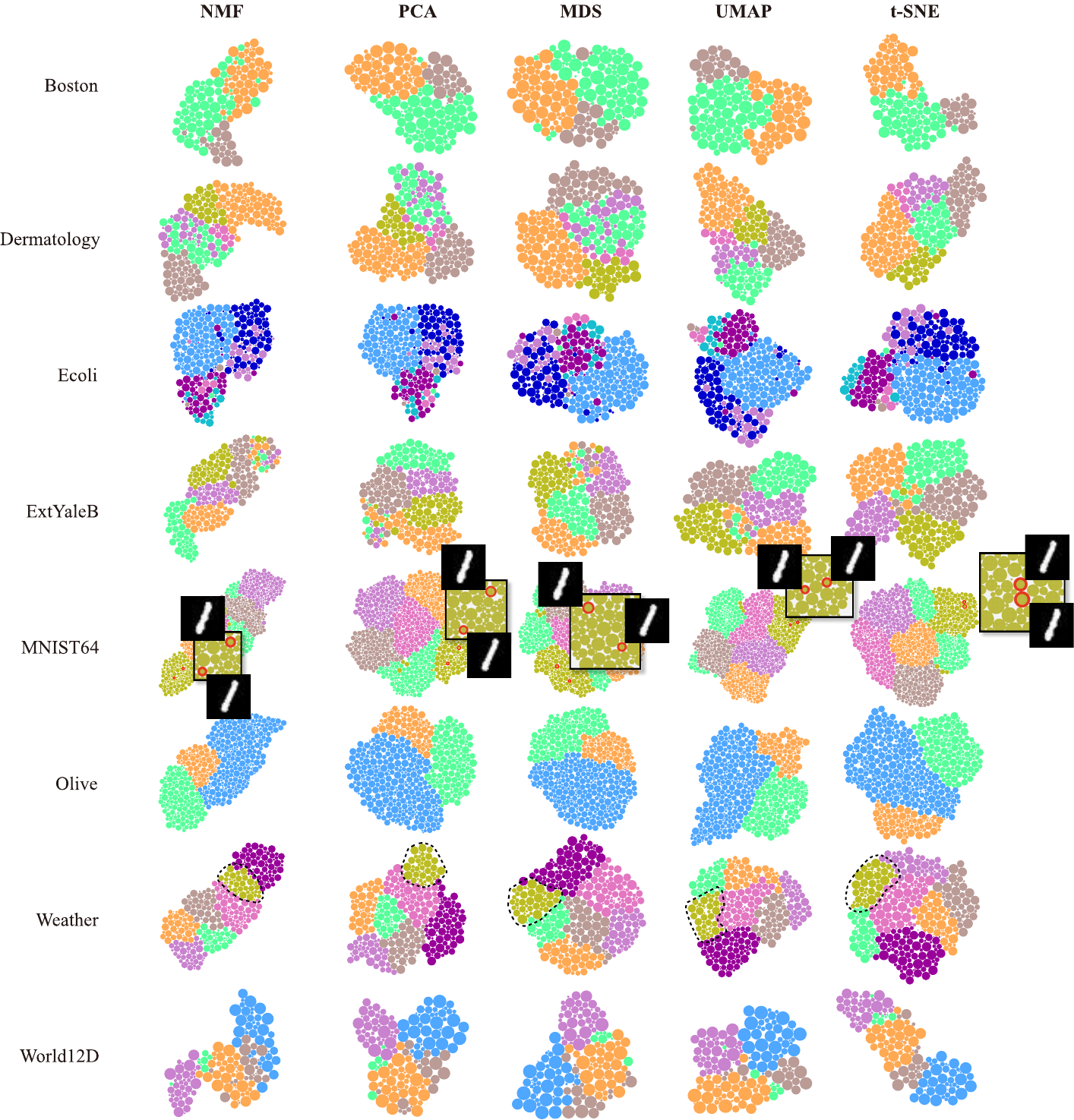}
\caption{Circle packing results for the eight datasets generated by alternative initialization methods where the circle radii have a small variance.}
\label{fig:sidebyside-init1}
\end{figure*}

\begin{figure*}[tp!]
\centering
\includegraphics[width=0.9\linewidth]{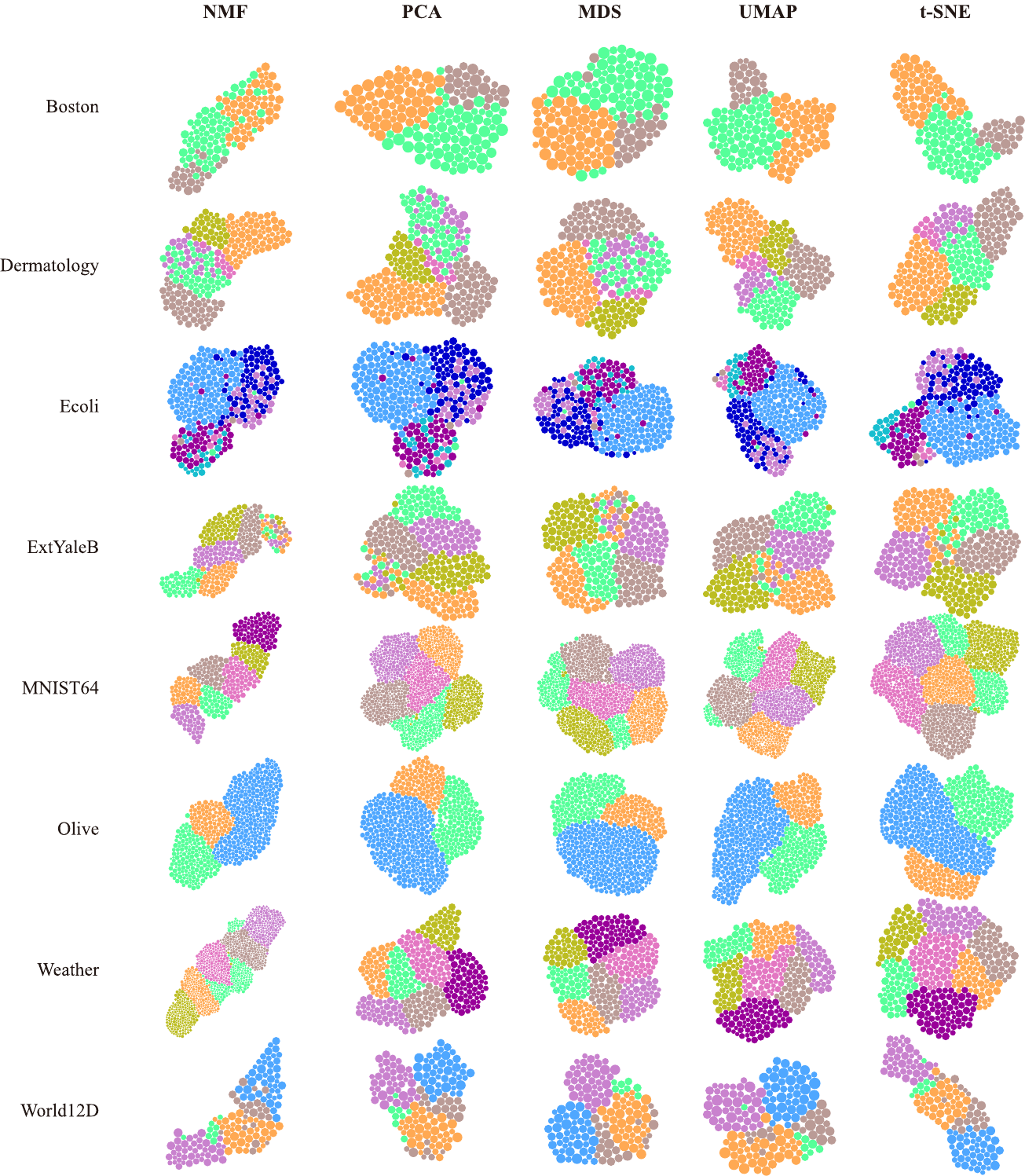}
\caption{Circle packing results for the eight datasets generated by alternative initialization methods where the circle radii have a large variance.}
\label{fig:sidebyside-init2}
\end{figure*}

\subsection*{Appendix C: Grid Search and Parameter Analysis}
Our optimization problem is formulated  in Eq.~\ref{eq:scalarized}.
{Since the parameters $\alpha$ and $\beta$, which balance the impact of the three terms,  affect the optimization result, we performed a grid search to determine the optimal parameters.}

Our measures involve neighborhood preservation ($NP_1$ and $NP_2$), compactness, and convexity, consistent with those in our quantitative evaluation.
We conducted a grid search to investigate the relationships between different measures and choices of weights $\alpha$ and $\beta$.
Here, we considered $\alpha \in [0.10, 0.20, 0.50, 1.00, 2.00, 5.00, 10.00]$ and $\beta \in [0.10, 0.20, 0.50, 1.00, 2.00, 5.00, 10.00]$.
The results are shown in \cref{fig:gridsearch-coarse}.
For neighborhood preservation, $NP_1$ and $NP_2$ decrease as $\alpha$ and $\beta$ increase: larger $\alpha$ and $\beta$ give less weight to neighborhood preservation $F_n$.
For compactness, the results are quite stable {(between 0.867 and 0.874)} when $\alpha$ and $\beta$ change due to use of sufficient iterations in the force-directed layout to guarantee convergence.
Convexity decreases as $\alpha$ increases or $\beta$ decreases, because larger $\alpha$ or smaller $\beta$ give less weight to $F_v$.
The observed trends in how the measures vary with changes in $\alpha$ and $\beta$ also demonstrate the effectiveness of adjusting optimization preferences by setting different values for $\alpha$ and $\beta$.

\begin{figure*}[t!]
\centering
\includegraphics[width=0.88\linewidth]{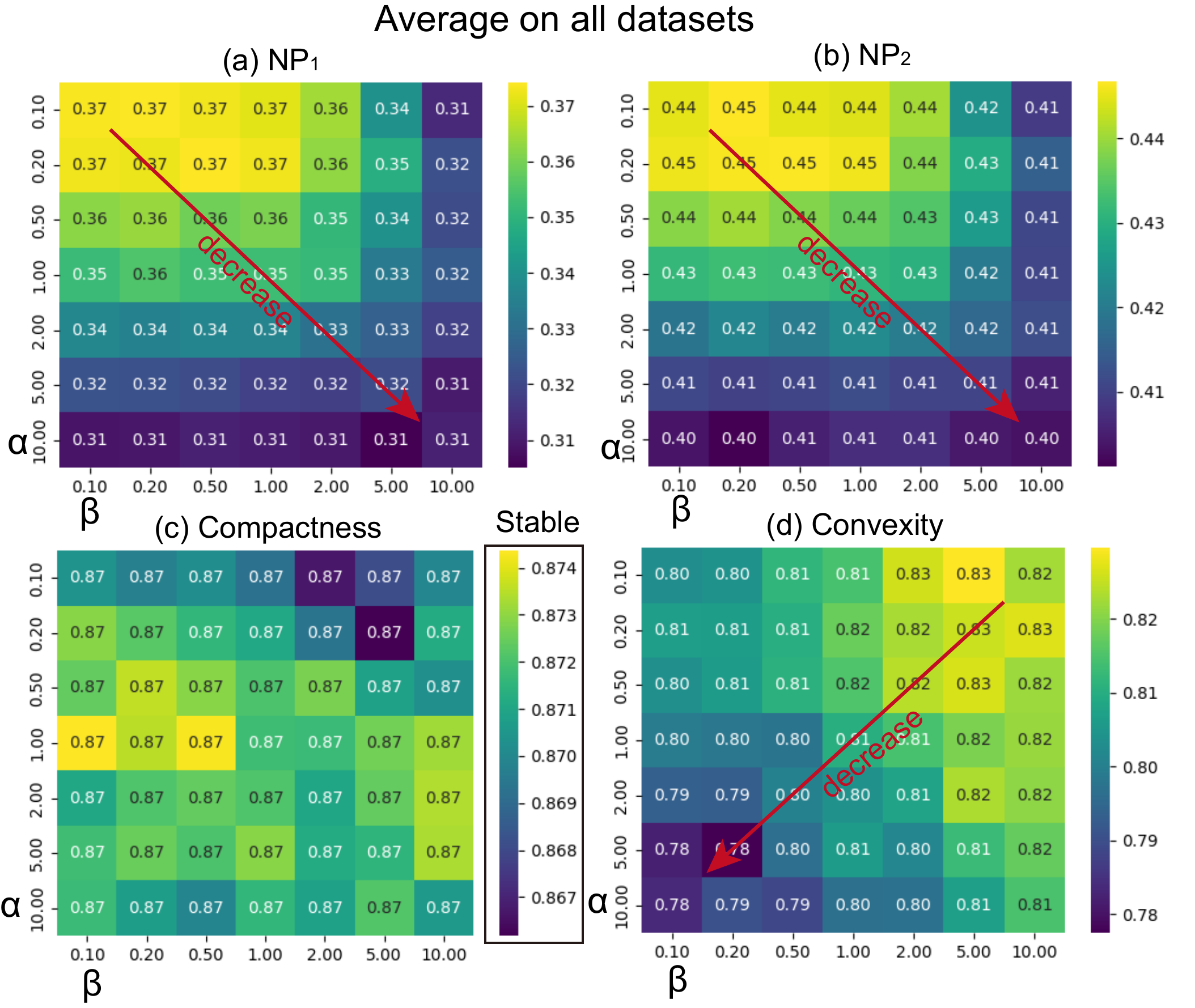}
\caption{Measure distributions with  varying $\alpha$ and $\beta$}
\label{fig:gridsearch-coarse}
\end{figure*}

To comprehensively compare the performance of our methods across different parameter settings, we designed a balanced index, which combines $NP_1$, $NP_2$, compactness, and convexity.
The balanced index is defined as the sum of all four measures.
We calculated the balanced index across all datasets, varying the parameters $\alpha$ and $\beta$ and taking the average.
{As illustrated in \cref{fig:gridsearch-sum}, the analysis revealed} a local maximum region with relatively high and stable balanced indexes {($\alpha\in[0.10,0.20,0.50]$ and $\beta\in[0.50,1.00,2.00]$),  with the peak occurring at} $\alpha=0.20$ and $\beta=1.00$.
{With this optimal parameter setting, the method delivers near-optimal performance across all evaluated dimensions: 99.5\% of the best $NP_1$, 99.8\% of the best $NP_2$, 99.6\% of the best compactness, and 98.7\% of the best convexity.}
Therefore, we chose it as the default parameter setting.

\begin{figure*}[t!]
\centering
\includegraphics[width=0.88\linewidth]{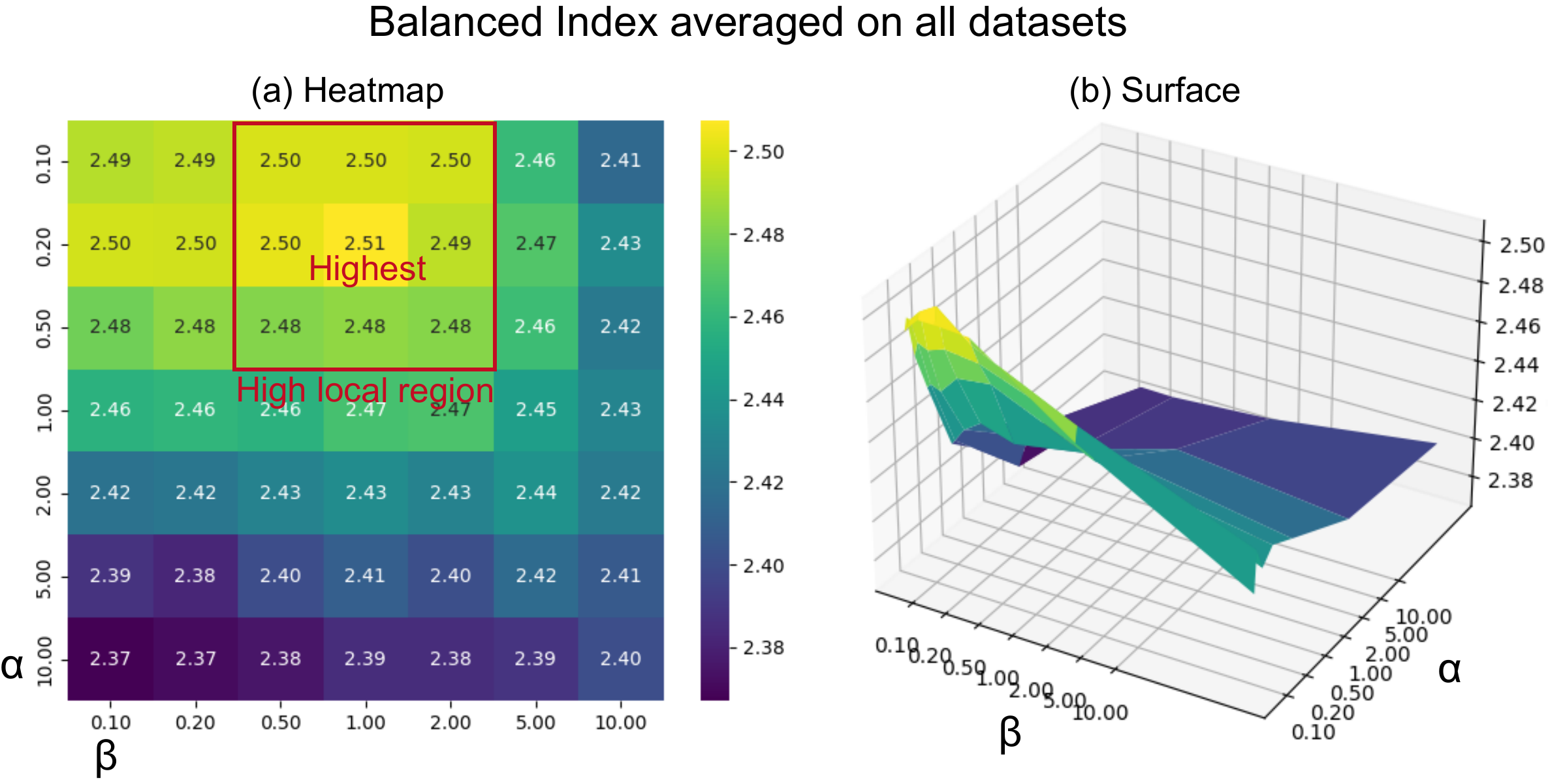}
\caption{Distribution of balanced index with varying $\alpha$ and $\beta$}
\label{fig:gridsearch-sum}
\end{figure*}

\subsection*{Appendix D: The Layout Results}
Figs.\ref{fig:sidebyside01} and~\ref{fig:sidebyside05} show the original t-SNE projection results (the first column) and circle packing generated by SimiFC, AEF, FD, and NCP (the remaining columns) for two  settings: large  and small radius variance, respectively.
Observing the shapes of clusters, it is easy to identify that SimiFC tends to generate arc-shaped clusters, and AEF tends to separate a cluster into multiple components.
Both  failed to provide a clear cluster structure, hindering analysis.
In contrast, both FD and NCP preserve the cluster structure well.
However, compared to the t-SNE projection result, FD often fails to preserve the relative positions of clusters.
Take the Boston dataset in \cref{fig:sidebyside01} as an example.
In the t-SNE projection result (\cref{fig:sidebyside01}A), the green cluster was placed in  between the orange cluster and the brown cluster, but FD dragged these three clusters together (\cref{fig:sidebyside01}B).
This would mislead users in understanding the similarity relationships between the three clusters.
Instead, NCP preserved the relative positions of clusters in the t-SNE projection well (\cref{fig:sidebyside01}C).
In addition, NCP usually forms smoother and more convex boundaries between the clusters (\eg, \cref{fig:sidebyside01}D versus \cref{fig:sidebyside01}E, and \cref{fig:sidebyside01}F versus \cref{fig:sidebyside01}G).
This further enhances the perception of cluster structures and facilitates data analysis.

\subsection*{Appendix E: Papers Used to Summarize the Design Criteria for \emph{NCP}}

\noindent Circle packing methods:
~\cite{wang2006circlepacking, goertler2018bubble, liu2016online, zhao2014fluxflow, zhao2015variational, yu2014content, rodrigues2023relaxed, samuel2013sedimentation, liu2016social, Blumenstein2017Livevis, yuan2023archexplorer, wang2020visual, viegas2013googleripples, wang2016unsupervised}.

\noindent Cluster-aware layout methods:
~\cite{zhou2023cluster, chen2021interactive, Rottmann2023MosaicSets, kehlbeck2021sp, efrat2014mapsets}.

\begin{figure*}[tp!]
\centering
\includegraphics[width=0.85\linewidth]{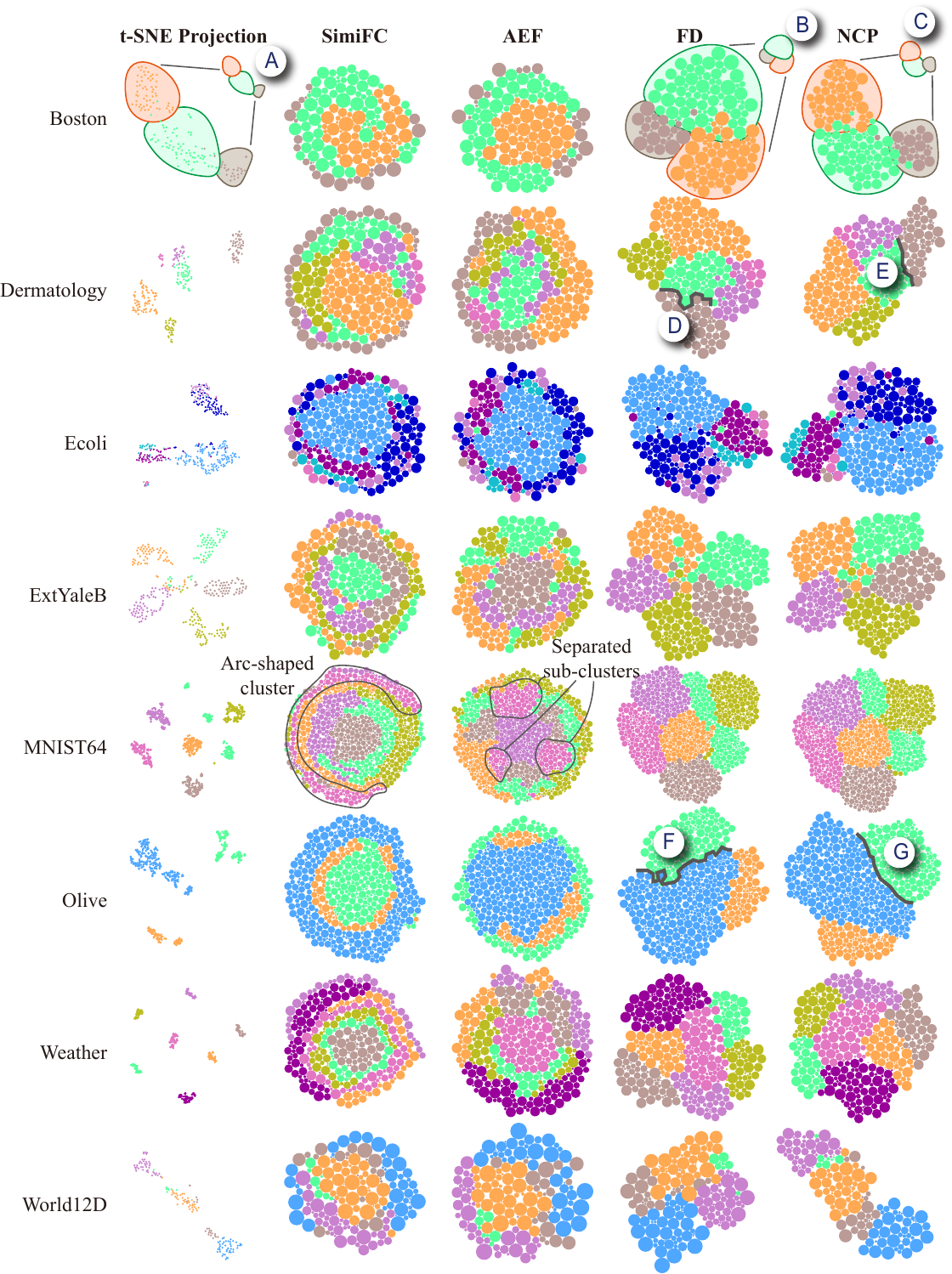}
\caption{Circle packing results for the eight datasets generated by SimiFC, {AEF}, FD, and NCP for circle radii with large variance.}
\label{fig:sidebyside01}
\end{figure*}

\begin{figure*}[tp!]
\centering
\includegraphics[width=0.85\linewidth]{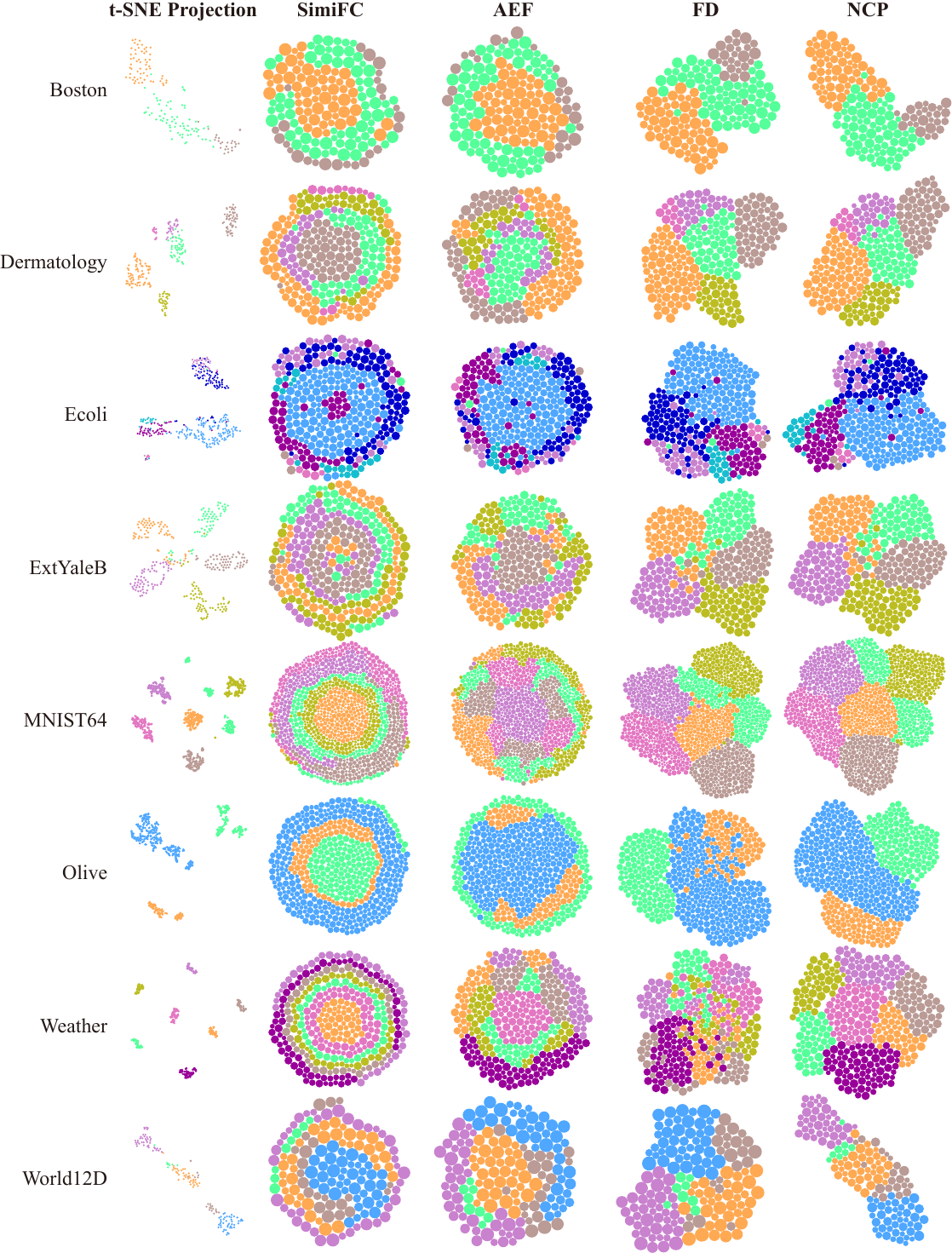}
\caption{Circle packing results of the eight datasets generated by SimiFC, {AEF}, FD, and NCP for circle radii with small variance.}
\label{fig:sidebyside05}
\end{figure*}

\subsection*{Availability of data and materials}

All data and materials are available on Github at \url{https://github.com/NCP-2024/NCP}.
In particular, they include datasets, source code,  a video, experimental results concerning the selection of projection methods and parameters, grid search analysis for algorithm parameters, detailed layout results, and the paper list used to summarize the design criteria for NCP.

\subsection*{Declaration of competing interest}

The authors have no competing interests to declare relevant to the content of this article.

\subsection*{Funding}

This work was supported by the National Natural Science Foundation of China (U21A20469, 61936002), and in part by Tsinghua University-China Telecom Wanwei Joint Research Center.

\subsection*{Authors' contributions}

\noindent Duan Li: implementation, algorithm design, writing---original draft, writing---review editing.

\noindent Jun Yuan: implementation, algorithm design, writing---original draft, writing---review editing.

\noindent Xinyuan Guo: implementation, algorithm design.

\noindent Xiting Wang: writing---review editing

\noindent Yang Liu: algorithm design, writing---review editing

\noindent Weikai Yang: writing---original draft, writing---review editing.

\noindent Shixia Liu: algorithm design, supervision, writing---original draft, writing---review editing, funding acquisition.

\subsection*{Acknowledgements}

The authors thank Dr. Changjian Chen, Dr. Qianwen Wang, Yukai Guo, Haoze Wang, and Jiangning Zhu for their valuable comments.

\bibliographystyle{CVMbib}
\bibliography{reference}

\subsection*{Author biographies}

\begin{biography}[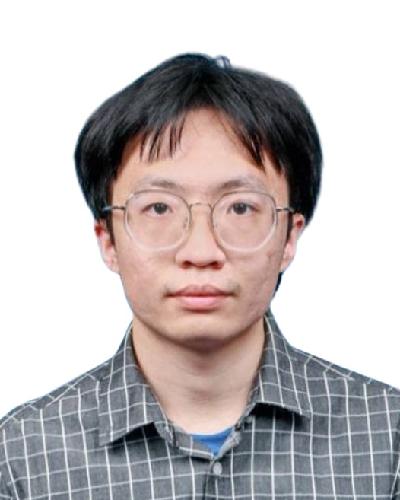]{Duan Li}
{
is a Ph.D. student at Tsinghua University, where he received his B.S. degree. His research interests lie in visual analytics. }
\end{biography}

\vspace{1cm}

\begin{biography}[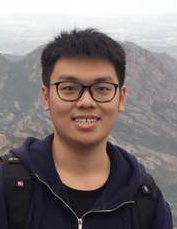]{Jun Yuan}
{
is currently a Ph.D. student at Tsinghua University, where he received his B.S. degree. His research interests lie in explainable artificial intelligence. 
}
\end{biography}

\vspace{1cm}

\begin{biography}[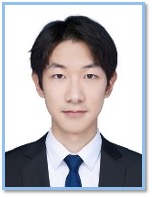]{Xinyuan Guo}
is currently a Ph.D. student at Tsinghua University. His research interests include visual analytics and animated transition design.
\end{biography}

\vspace{1cm}

\begin{biography}[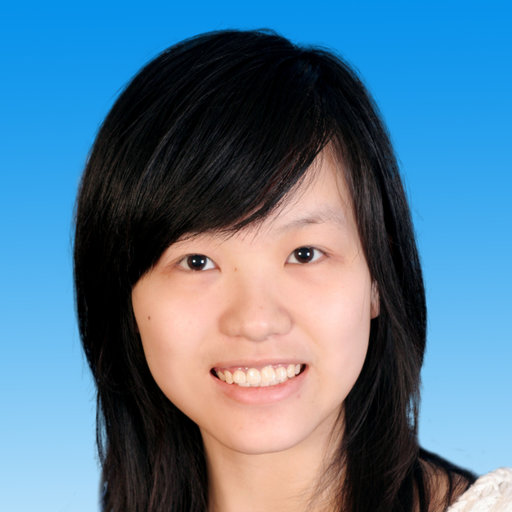]{Xiting Wang}
{
is an assistant professor at Renmin University of China. Her research interests include explainable machine learning and visual text analytics. She has  academic papers in outlets such as KDD,  IEEE Transactions on Knowledge and Data Engineering, AAAI, IJCAI,  IEEE Transactions on Visualization and Computer Graphics and VAST.
}
\end{biography}

\vspace{1cm}

\begin{biography}[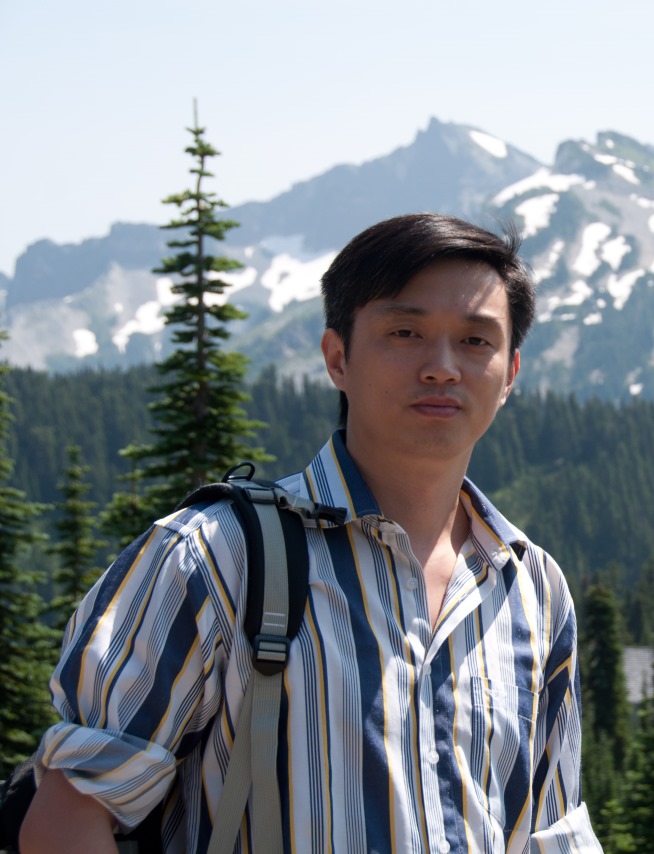]{Yang Liu}
{
is a principal researcher at Microsoft Research Asia. He received his Ph.D. degree from The University of Hong Kong, and master's and bachelor's degrees from The University of Science and Technology of China. His recent research focuses on geometric computation and learning-based geometry processing and generation. He is an associate editor of Transactions on Visualization and Computer Graphics.
}
\end{biography}

\vspace{1cm}

\begin{biography}[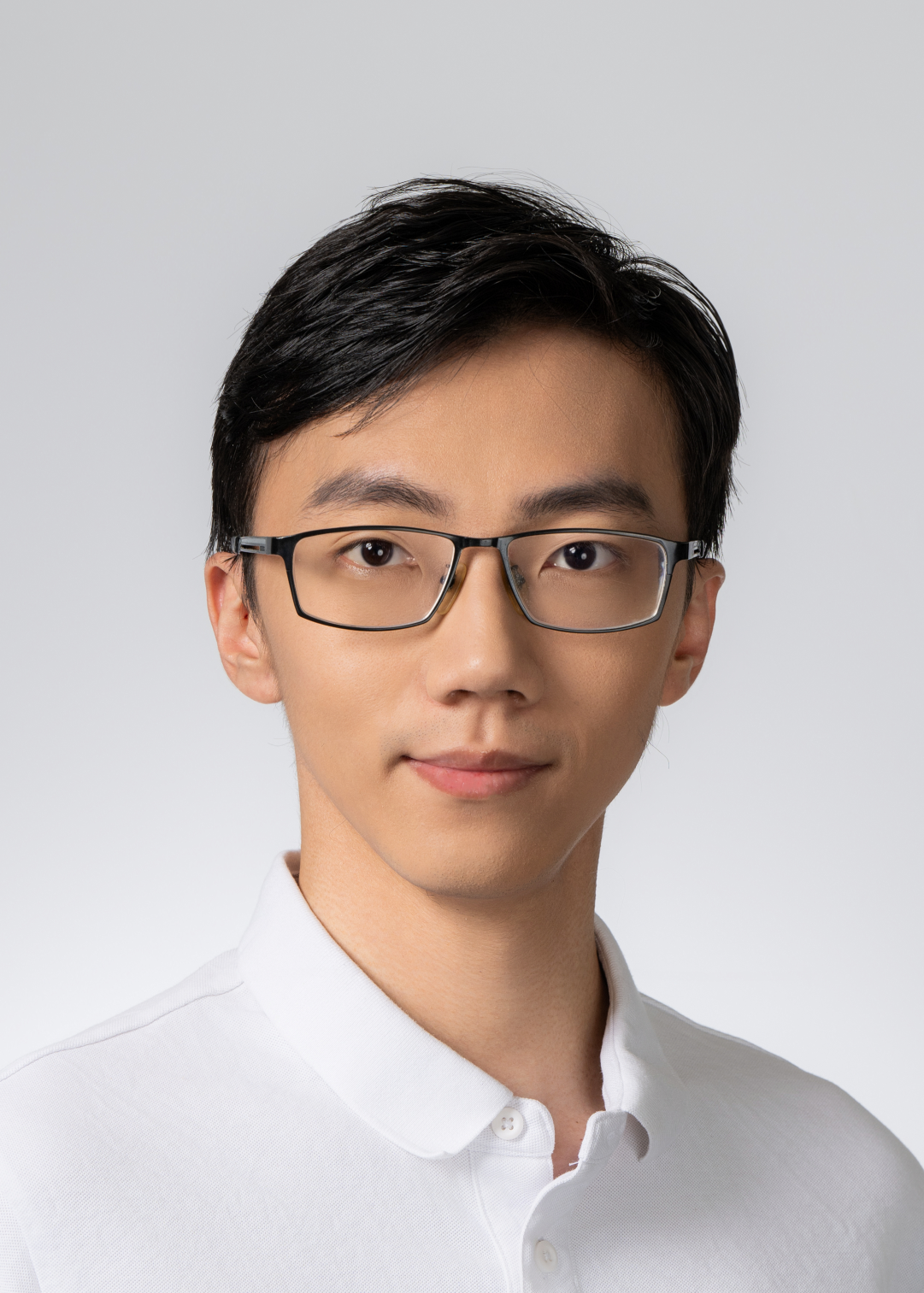]{Weikai Yang}
{
is an assistant professor in Hong Kong University of Science and Technology (Guangzhou). His research interests lie in visual analytics, machine learning, and data quality improvement. He received  B.S. and  Ph.D degrees from Tsinghua University.
}
\end{biography}

\vspace{1cm}

\begin{biography}[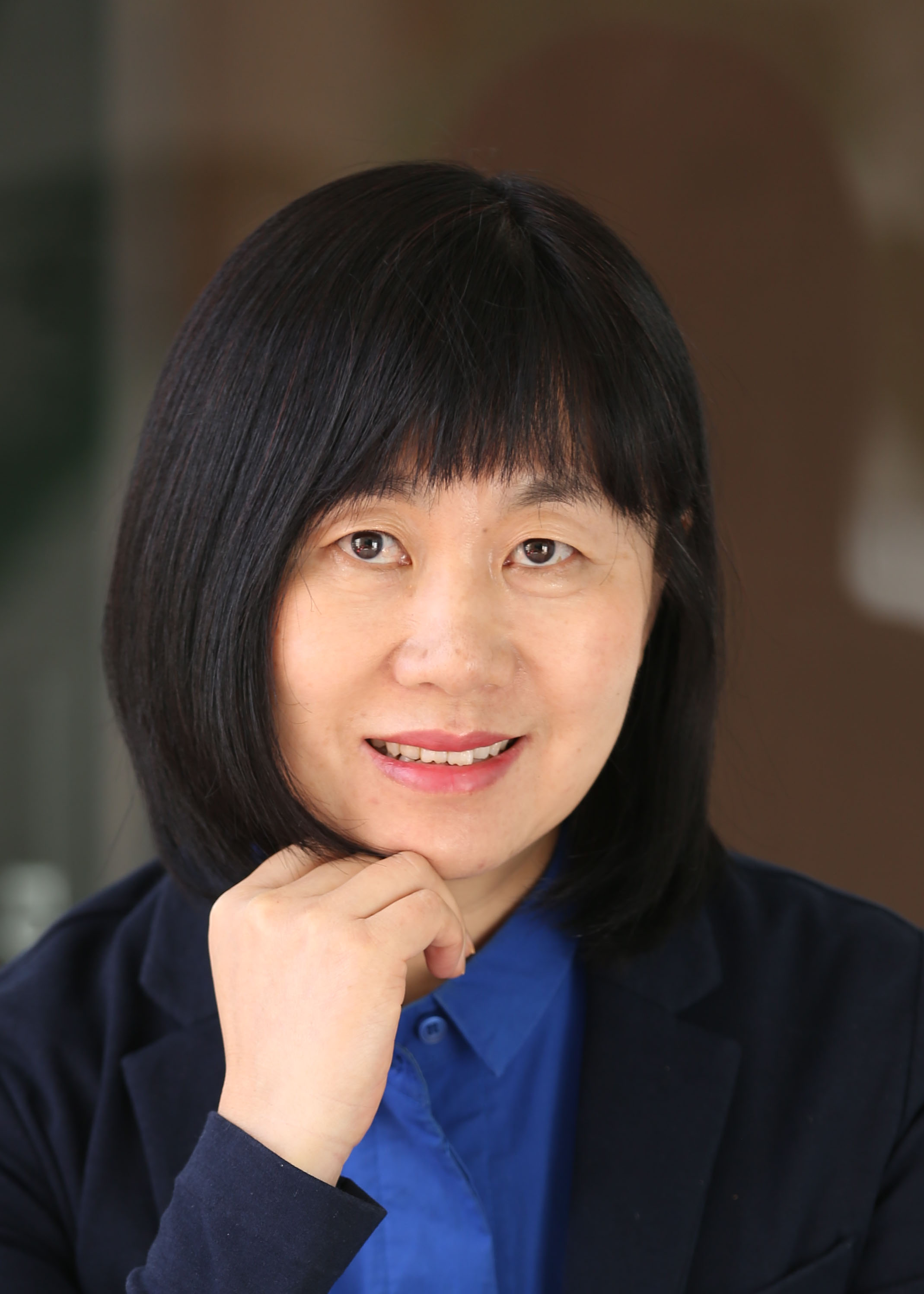]{Shixia Liu}
{
is a professor at Tsinghua University. Her research interests include visual text analytics, visual social analytics, interactive machine learning, and text mining. She has worked as a research staff member at IBM China Research Lab and was a lead researcher at Microsoft Research Asia.
She received  B.S. and M.S. degrees from Harbin Institute of Technology, and a Ph.D. from Tsinghua University.
She is a fellow of IEEE and an  associate editor-in-chief of Transactions on Visualization and Computer Graphics.
}
\end{biography}

\vspace{2cm}

\subsection*{Graphical abstract}
\begin{figure*}[!htb]
\centering
  \includegraphics[width=\linewidth]{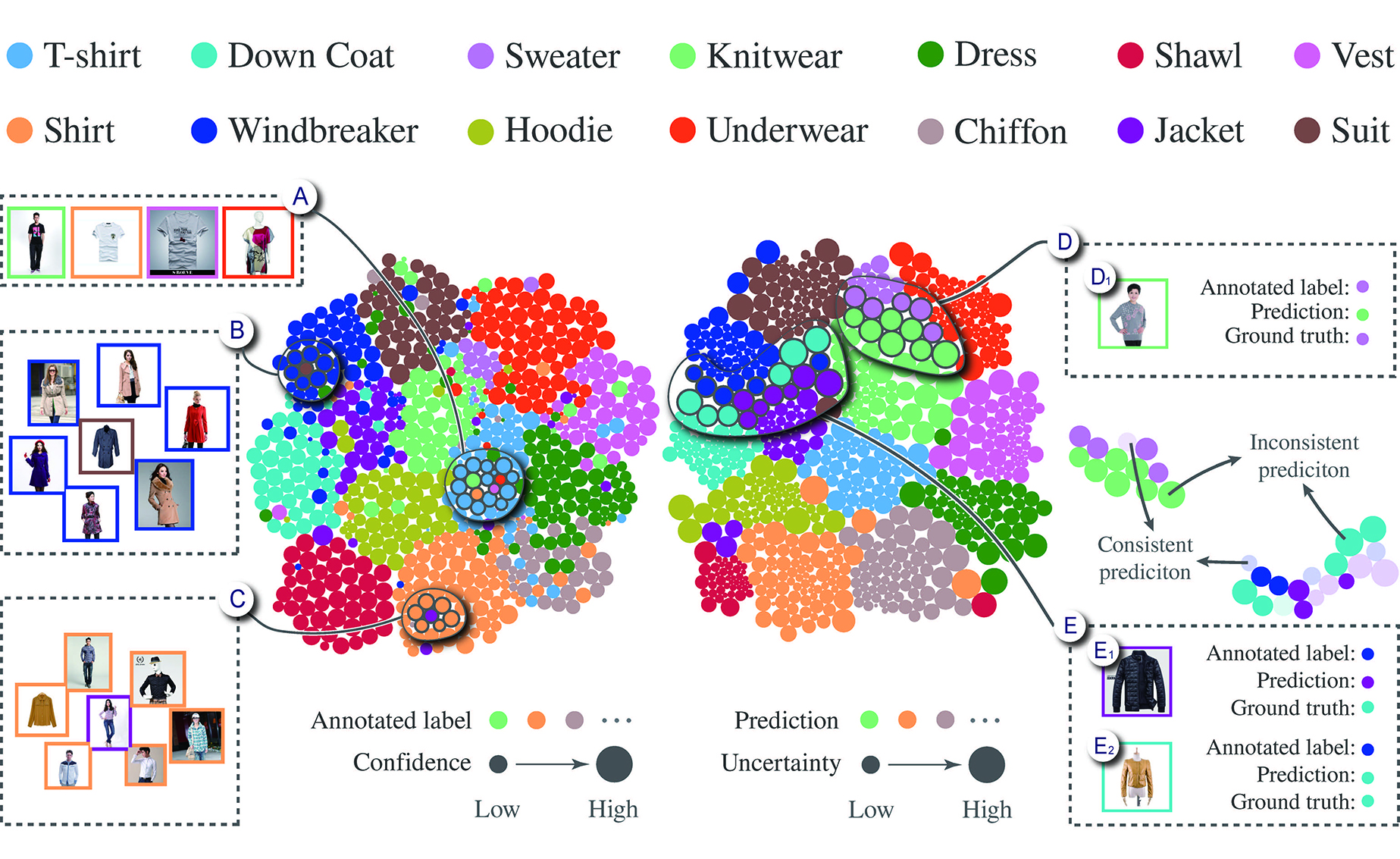}
  \label{fig:graphicalabstract}
\end{figure*}

\end{document}